\documentclass[useAMS,usenatbib,usegraphicx]{mn2e}
\begin{document}
%
%
\title[Star formation in NGC~2974]{Star formation and figure rotation
  in the early-type galaxy NGC~2974}
\author[H.\ Jeong et al.]{Hyunjin Jeong,$^{1,2}$\thanks{E-mail:
hyunjin@galaxy.yonsei.ac.kr} Martin Bureau,$^2$ Sukyoung K.\
Yi,$^{1,2}$ Davor Krajnovi\'{c}$^2$, and \newauthor Roger L.\ Davies$^2$\\
$^1$Department of Astronomy, Yonsei University, Seoul 120-749,
    Korea\\
$^2$Sub-Department of Astrophysics, University of Oxford, Denys
Wilkinson Building, Keble Road, Oxford OX1 3RH}
\maketitle
%
%
\begin{abstract}
We present {\it Galaxy Evolution Explorer} ({\it GALEX}) far (FUV) and
near (NUV) ultraviolet imaging of the nearby early-type galaxy
NGC~2974, along with complementary ground-based optical imaging. In
the ultraviolet, the galaxy reveals a central spheroid-like component
and a newly discovered complete outer ring of radius $6.2$~kpc, with
suggestions of another partial ring at an even larger radius. Blue
FUV$-$NUV and UV$-$optical colours are observed in the centre of the
galaxy and from the outer ring outward, suggesting young stellar
populations ($\la1$~Gyr) and recent star formation in both
locations. This is supported by a simple stellar population model
which assumes two bursts of star formation, allowing us to constrain
the age, mass fraction and surface mass density of the young component
pixel by pixel. Overall, the mass fraction of the young component
appears to be just under $1$~per cent (lower limit, uncorrected for
dust extinction). The additional presence of a nuclear and an inner
ring (radii $1.4$ and $2.9$~kpc, respectively), as traced by
[O\,{\small III}] emission, suggests ring formation through
resonances. All three rings are consistent with a single pattern speed
of $78\pm6$ km~s$^{-1}$~kpc$^{-1}$, typical of S0 galaxies and only
marginally slower than expected for a fast bar if traced by a small
observed surface brightness plateau.  This thus suggests that star
formation and morphological evolution in NGC~2974 at the present epoch
are primarily driven by a rotating asymmetry (probably a large-scale
bar), despite the standard classification of NGC~2974 as an E4
elliptical.
\end{abstract}
%
%
\section{INTRODUCTION}
\label{sec:intro}
Early-type galaxies are traditionally viewed as dynamically simple
stellar systems with homogeneous stellar populations
\citep[e.g.][]{g77}. However, building on earlier work (e.g.\
\citealt{zf91} and references therein), the {\tt SAURON} survey
\citep{betal01,zetal02} of the two-dimensional ionised-gas/stellar
kinematics and stellar populations of nearby early-type galaxies has
recently revealed kinematically-decoupled components and substantial
metallicity and/or age gradients in many objects
\citep[e.g.][]{eetal04,setal06,kuetal06}. Early-type galaxies are thus
likely to have had complex and varied formation histories. Such
properties are often taken as evidence against the classical
monolithic collapse model \citep*[e.g.][]{els62,la74}, in which
elliptical galaxies are old systems that formed early in highly
efficient starbursts and have evolved passively since.

Indeed, while standard optical imaging shows that most early-type
galaxies have red colours and old stellar populations, deep imaging
surveys have shown that many possess shells and tidal features
\citep*[e.g.][]{ss92}, telltale signs of a turbulent past. Colour
gradients also indicate that early-type galaxies have composite
stellar populations \citep[e.g.][]{pe90}, and some nearby early-type
galaxies have signatures of ongoing or recent star formation
\citep*[RSF; e.g.\ NGC~2865;][]{hcb99,retal05}. Spectroscopic studies
further suggest that many elliptical galaxies have intermediate-age
stars \citep[e.g.][]{oco80,ba87,tetal00}. The spread in the
luminosity-weighted age \citep[e.g.][]{tetal00} is evidence for
RSF. The so-called E~+~A galaxies \citep{dg83} possess characteristic
spectral features defined by strong hydrogen Balmer absorption lines
(H$\delta$, H$\gamma$, H$\beta$), indicating the presence of a young
stellar population, but have negligible [O\,{\small II}] $\lambda$3727
emission. These galaxies are believed to be `post-starburst' systems,
i.e.\ they have recently undergone a burst of star
formation. \citet{fetal04} also suggested that a small fraction of low
redshift elliptical galaxies have unambiguous signatures of active
star formation with rates as high as those in spirals.

Far ultraviolet (FUV) radiation was first discovered in early-type
galaxies by the {\it Orbiting Astronomical Observatory-2} in 1969
\citep*{cwp72}, but it is now generally recognized that UV emission is
a ubiquitous phenomenon in these objects. Although star formation (SF)
can also be measured indirectly at other wavelengths, particularly in
the infrared \citep[e.g.][]{hetal86,metal97} and through H$\alpha$
emission \citep*[e.g.][]{ke83,hbi03}, UV light is very sensitive to
stellar populations younger than a few hundred megayears and the UV
flux can give important and direct information about the star
formation rate (e.g.\ \citealt{hbi03}; \citealt*{betal05}). The FUV is
sensitive to the UV upturn flux originating from old hot
helium-burning horizontal branch (HB) stars (see, e.g.,
\citealt*{ydo97} and \citealt{oco99} for reviews of this issue and the
UV upturn phenomenon), hence SF studies usually focus on the near
ultraviolet (NUV) passband. \citet{fs00} studied young stars in the
cluster Abell~851 ($z=0.41$) using NUV$-$optical colours and suggested
that, for some of the early-type galaxies, the stellar mass fraction
in young stars is higher than $10$~per cent. More recently,
\citet{yetal05} investigated the NUV colour-magnitude relation of
early-type galaxies classified by the Sloan Digital Sky Survey. They
concluded that roughly $15$~per cent of nearby bright ($M_r<-22$)
early-type galaxies show signs of recent ($\la1$~Gyr) star formation
(at the $1-2$~per cent level by stellar mass).

Here, we focus on the nearby early-type galaxy NGC~2974, classified as
E4 in the RC3 \citep{vetal91} but possessing many features (mainly
dynamical) more reminiscent of lenticular galaxies \citep[see,
e.g.,][]{cm94}. NGC~2974 has in fact been widely studied because of
its unusual physical properties, including a dust lane containing
ionised gas \citep[e.g.][]{k89} and substantial neutral hydrogen
(H\,{\small I}) consistent with a rotating disc aligned with the
optical isophotes \citep[e.g.][]{kjgkg88}. \citet{betal93} also
suggested that it has a complex lens-like feature. Recently, using
data from the {\tt TIGER} integral-field spectrograph and {\it Hubble
Space Telescope}'s ({\it HST}) Wide Field and Planetary Camera~2
(WFPC2), \citet*{egf03} discovered a gaseous two-arm spiral structure
in the central $500$~pc, suggesting the presence of a nuclear bar. The
larger scale {\tt SAURON} integral-field observations of
\citet{kcemz05} reveal an apparent nuclear bar, nuclear ring, larger
scale spiral arms and a possible inner ring in the ionised gas map
(see Section~\ref{sec:rings}), which suggest the presence of a
large-scale bar as well.

In this paper, we present and discuss new UV imaging observations of
NGC~2974 obtained with {\it Galaxy Evolution Explorer} ({\it
GALEX}). In Section~\ref{sec:obs}, we describe the {\it GALEX}
observations and data reduction, supporting optical imaging, and the
main results from UV$-$optical surface photometry. Recent star
formation in NGC~2974 is quantified and discussed in
Section~\ref{sec:spop_sf}, and evidence for a rotating pattern, mainly
based on the discovery of a star-forming outer ring, is discussed in
Section~\ref{sec:fig_rot}. We summarize our results and discuss their
implications briefly in Section~\ref{sec:summary}.
%
%
\section{OBSERVATIONS AND MAIN RESULTS}
\label{sec:obs}
\subsection{UV observations and data analysis}
\label{sec:uv}
We observed NGC~2974 with the Medium imaging mode of {\it GALEX} on
2005 February 19, part of a larger survey of the galaxy sample from
the {\tt SAURON} project \citep[see][]{zetal02}. A NASA small explorer
mission, {\it GALEX} is performing the first ultraviolet sky survey
from space. It consists of a $50$-cm UV$-$optimized modified
Ritchey-Chr\'{e}tien telescope, with a circular field-of-view (FOV) of
$1\fdg2$ diameter. The {\it GALEX} instruments and mission are
described fully in \citet{maetal05} and \citet{moetal05}. Exposure
times for NGC~2974 were $1477$~s in both FUV ($1350$--$1750$~\AA) and
NUV ($1750$--$2750$~\AA). Although the images are delivered
pre-processed, we performed our own sky subtraction by measuring the
sky level in source-free regions of the images. The spatial resolution
of the images is approximately $4\farcs5$ and $6\farcs0$ FWHM in FUV
and NUV, respectively, sampled with $1\farcs5\times1\farcs5$
pixels. We convolve the FUV data to the spatial resolution of the NUV
observations to avoid spurious colour gradients in the inner parts.

We have performed surface photometry of NGC~2974 by measuring the
surface brightness along elliptical annuli in the standard manner,
using the ELLIPSE task within the STSDAS ISOPHOTE package in IRAF
(Image Reduction and Analysis Facility). The centre of the isophotes
was fixed to the centre of the light distribution and the position
angle (PA), ellipticity ($\epsilon$) and surface brightness ($\mu$)
were fitted as a function of radius, the latter increasing
logarithmically to compensate for the rapid surface brightness
decrease. The ellipses were fitted to the NUV image only, which has a
far superior signal-to-noise ratio (S/N) at all radii, and were then
imposed to the FUV image, so that meaningful colours can be
derived. The fit fails at large radii due to low fluxes, so the
position angle and ellipticity are kept fixed at those radii and only
the surface brightness is measured. The necessary photometric
zero-points were taken from \citet{moetal05}. Our UV surface
brightness profiles extend to $2\arcmin$ ($12.5$~kpc for an assumed
distance of $21.5$~Mpc; \citealt{tdbaflmm01}) and reach depths of
$29.0$ and $28.6$ AB mag~arcsec~$^{-2}$ in FUV and NUV,
respectively. However, the nominal surface brightness limit for the
$1477$~s exposures used in our observations is roughly $28$~AB
mag~arcsec~$^{-2}$ in both bands \citep{maetal05}.  The difficulty of
the sky subtraction is the main limiting factor.
%
%

\begin{figure*}
\begin{center}
\includegraphics[width=8cm]{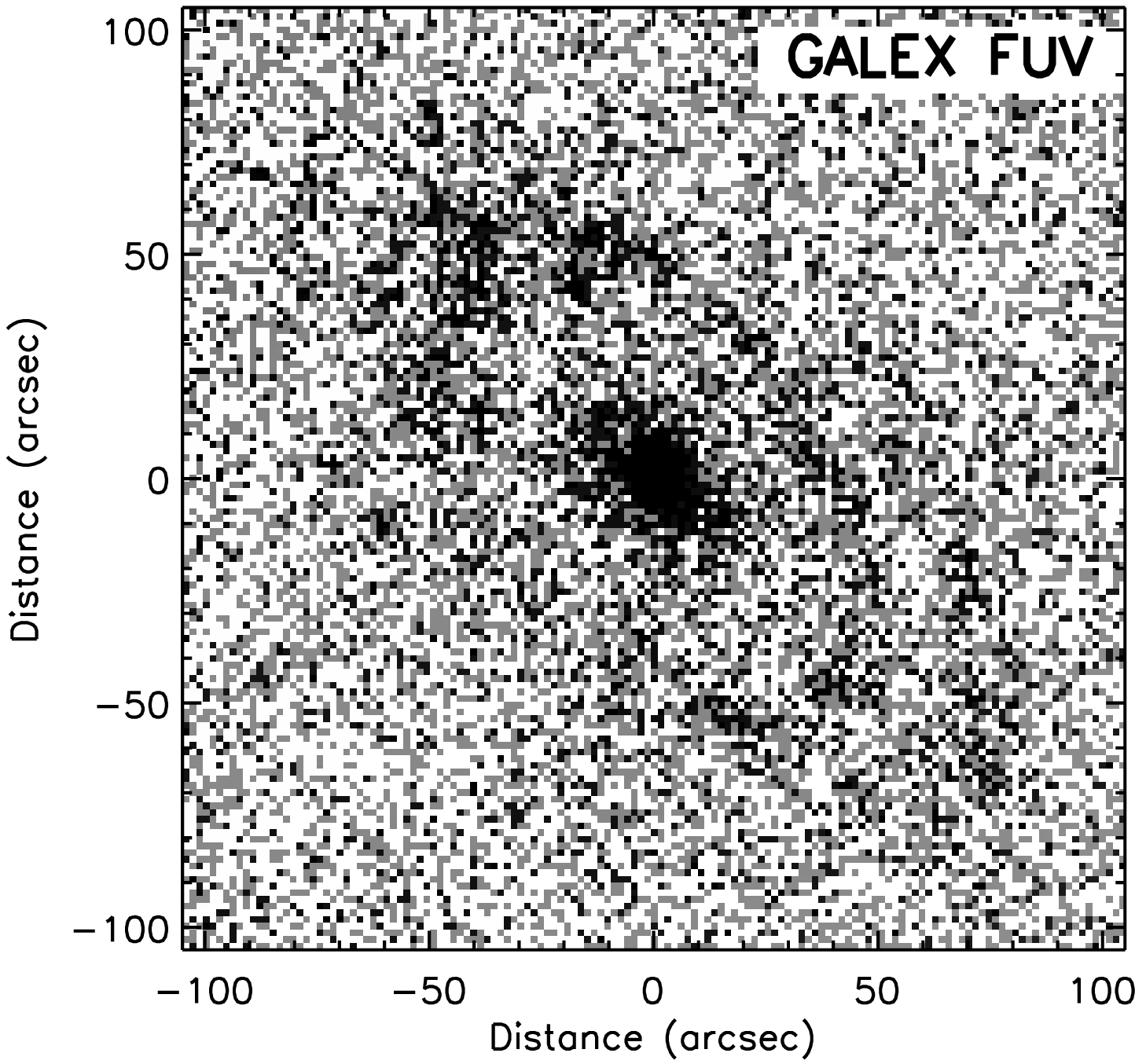}
\includegraphics[width=8cm]{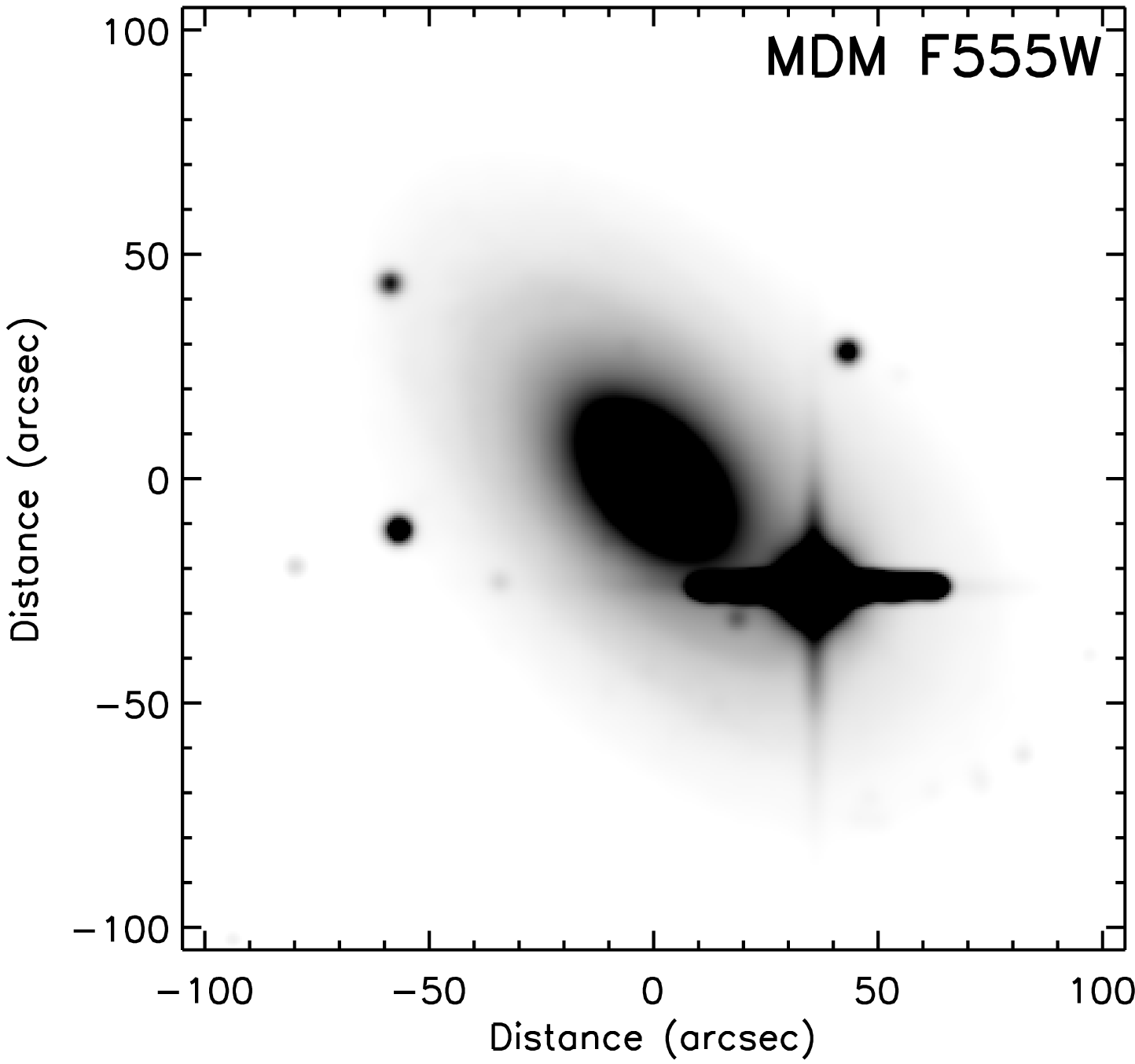}\\
\vspace*{3mm}
\includegraphics[width=8cm]{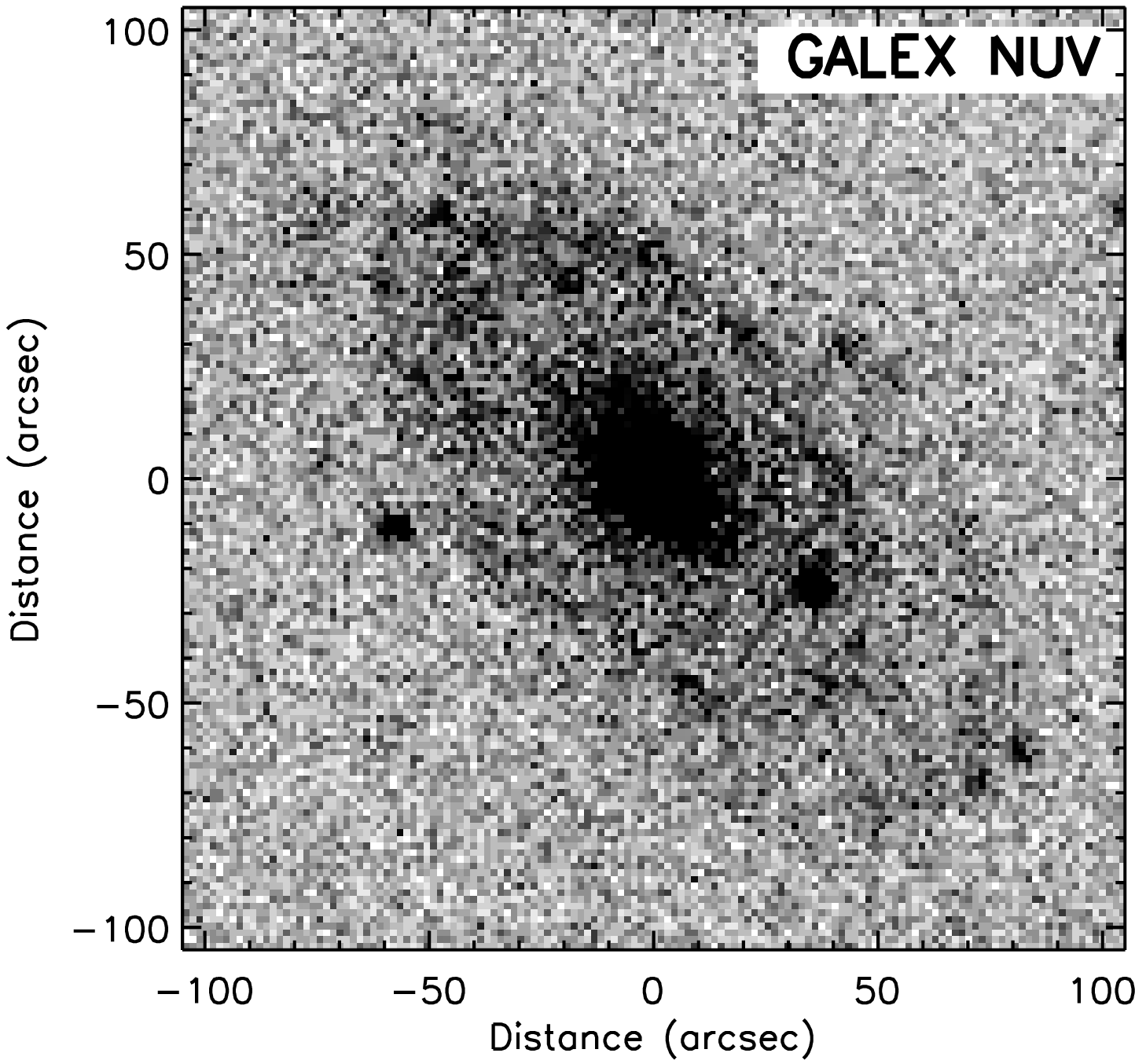}
\vspace*{3mm}
\includegraphics[width=8cm]{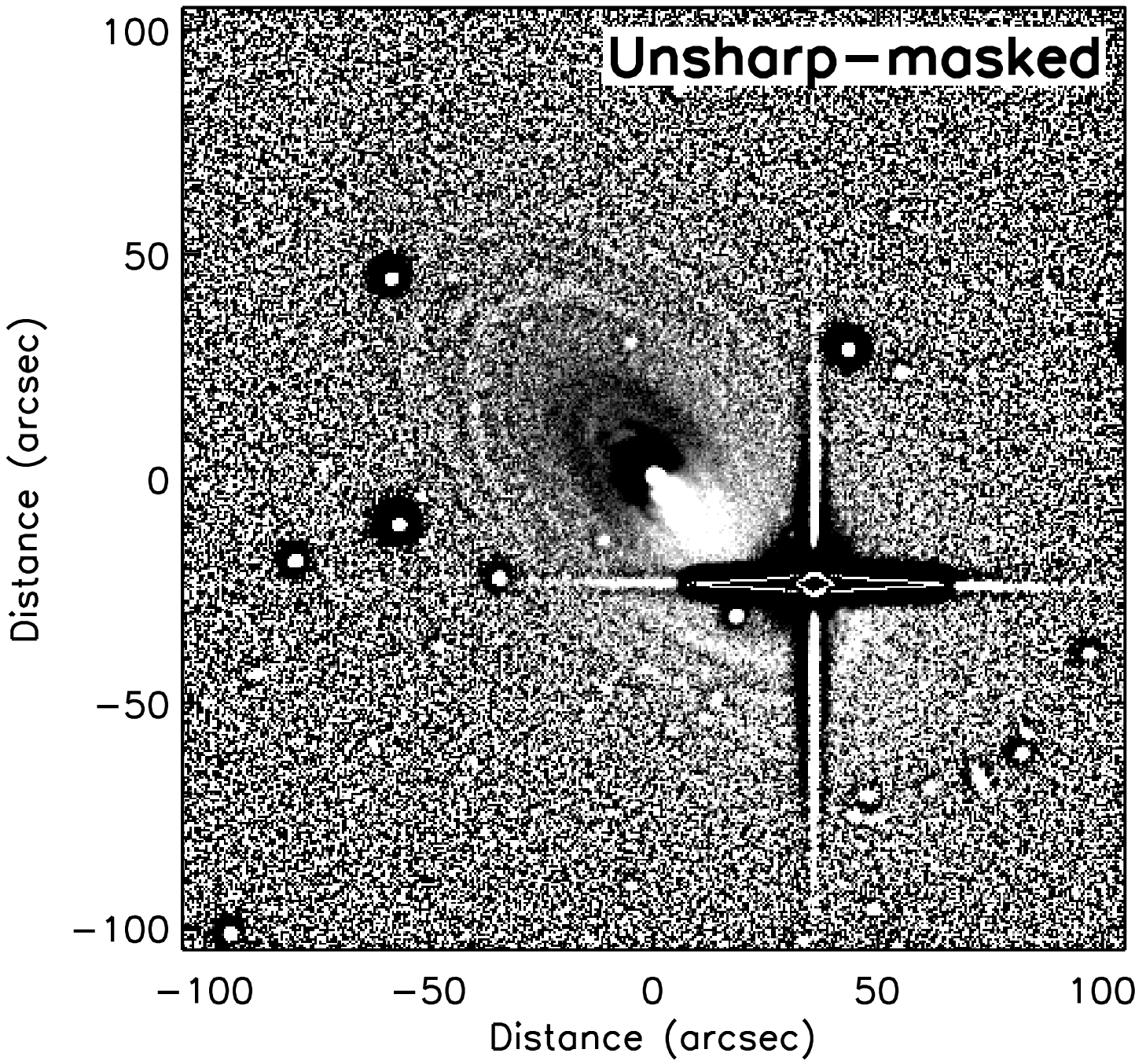}
\caption{{\it GALEX} UV and MDM optical images of the central
  $100\arcsec\times100\arcsec$ of NGC~2974. {\em Top-left:} FUV
  image. {\em Bottom-left:} NUV image. {\em Top-right:} F555W
  image. {\em Bottom-right:} Unsharp-masked F555W image. Note the
  bright foreground star to the South-West.}
\label{fig:images}
\end{center}
\end{figure*}
\subsection{Optical observations and data analysis}
\label{sec:optical}
Ground-based optical imaging observations in the {\it HST} filter
F555W (similar to Johnson $V$) were obtained with the MDM Observatory
1.3-m McGraw-Hill Telescope on 2003 March 26, again part of a larger
survey targeting the whole {\tt SAURON} galaxy sample.  The MDM
observations are described in detail in Falc\'{o}n-Barroso et al.\
(2007, in prep.) and were reduced in the standard manner in IRAF. The
FOV of the MDM images is $17\farcm3\times17\farcm3$ with
$0\farcs508\times0\farcs508$ pixels, allowing for accurate sky
subtraction and proper sampling of the seeing. The seeing for the
NGC~2974 observations was $1\farcs2$, but the MDM data were also
convolved to the resolution of the NUV data.

Surface photometry along ellipses was carried out for the MDM data in
the same manner as for the {\it GALEX} data. The NUV ellipses were
imposed on the MDM data and only the surface brightness was measured
for each ellipse, allowing accurate UV$-$optical colours to be
derived. An identical procedure was applied to {\it HST} WFPC2 images
in the F555W filter, obtained from the Space Telescope European
Coordinating Facility (ST-ECF) {\it HST} archive. The photometric
zero-point of the MDM data was derived by scaling the MDM surface
brightness profile to that of the {\it HST} data, allowing for an
arbitrary background level (the limited FOV of WFPC2 does not permit
accurate sky subtraction). The MDM data necessary to photometrically
calibrate the NGC~2974 imaging are available, but an HST-based
calibration is both more accurate and more simple. Given the limited
spatial resolution of {\it GALEX}, the {\it HST} data were used only
to calibrate the MDM photometry. The MDM profiles extend to much
greater radii than the {\it GALEX} profiles, but only the overlapping
region is discussed here.
\subsection{Results}
\label{sec:results}
Figure~\ref{fig:images} shows the {\it GALEX} FUV, {\it GALEX} NUV,
MDM F555W, and MDM F555W unsharp-masked images. The optical image has the smooth appearance
characteristic of elliptical galaxies, but the UV images reveal at
least two components. First, both FUV and NUV images show a largely
featureless central component with a rapidly decreasing surface
brightness, which can probably be identified as a spheroid. Second,
most distinct in the FUV, a complete ring is detected in the outer
parts of the galaxy, at a radius of $\approx60\arcsec$. The UV
emission along this ring is patchy and it probably lies in a disc (see
below). A partial ring is also suggested at a larger radius
($R\approx90\arcsec$), wrapping over roughly $90\degr$. Both rings are
detected here for the first time and are primarily visible in the
UV. However, a posteriori unsharp-masking of the MDM F555W image at
full resolution does reveal hints of multiple narrow rings or tightly
wound spirals arms at the right radii, more prominent in the
North-East (see Fig.~\ref{fig:images}). Aperture photometry on the FUV
image within the outer ring ($R\la60\arcsec$) yields a UV luminosity
of $\approx4.4~\times10^{8}$~$L_{\odot}$. The luminosity in the outer
ring itself ($50\arcsec\la R\la70\arcsec$) is
$\approx1.2~\times10^{8}$~$L_{\odot}$.

\begin{figure}
\begin{center}
\includegraphics[width=6.5cm]{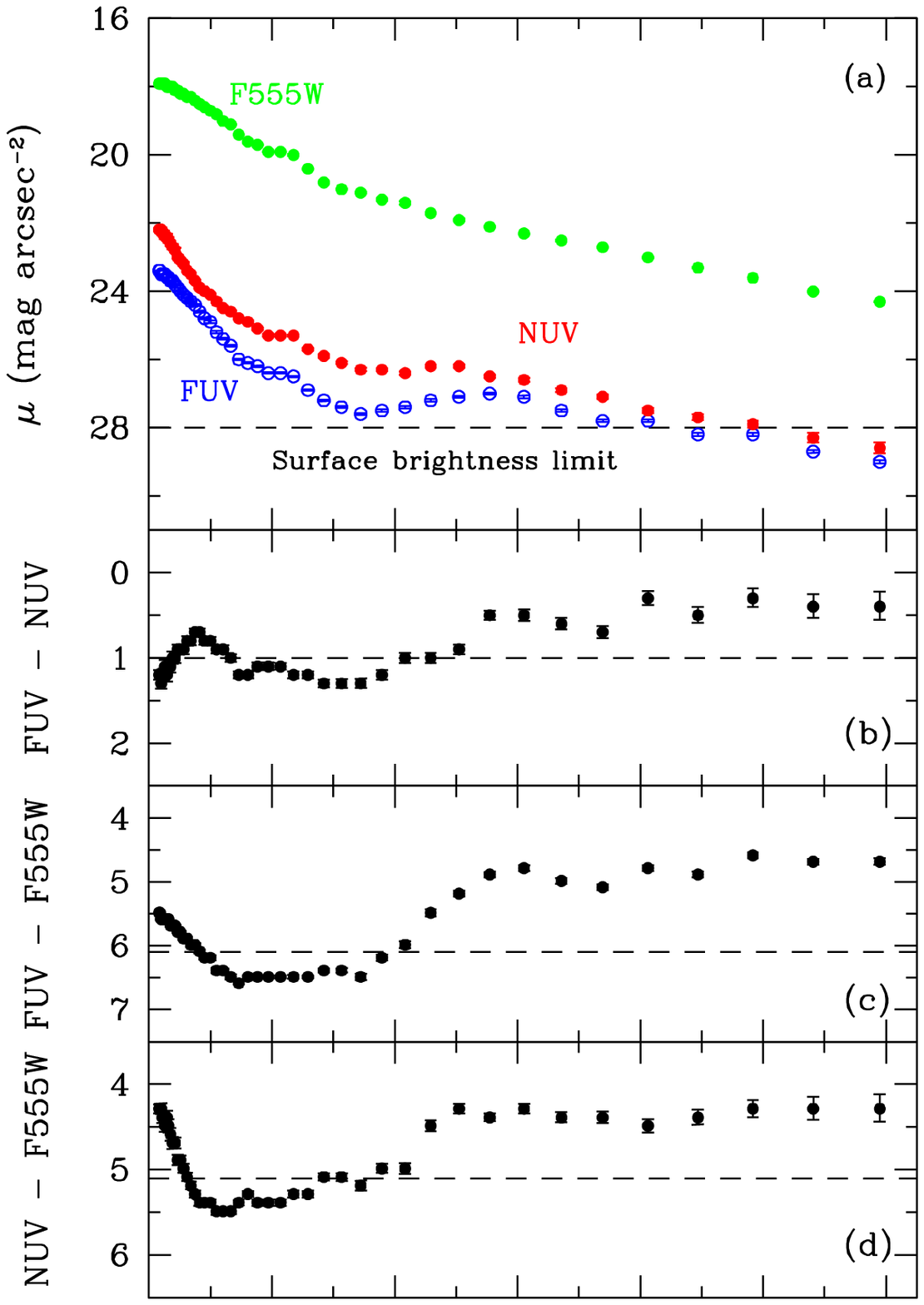}\\
\includegraphics[width=6.5cm]{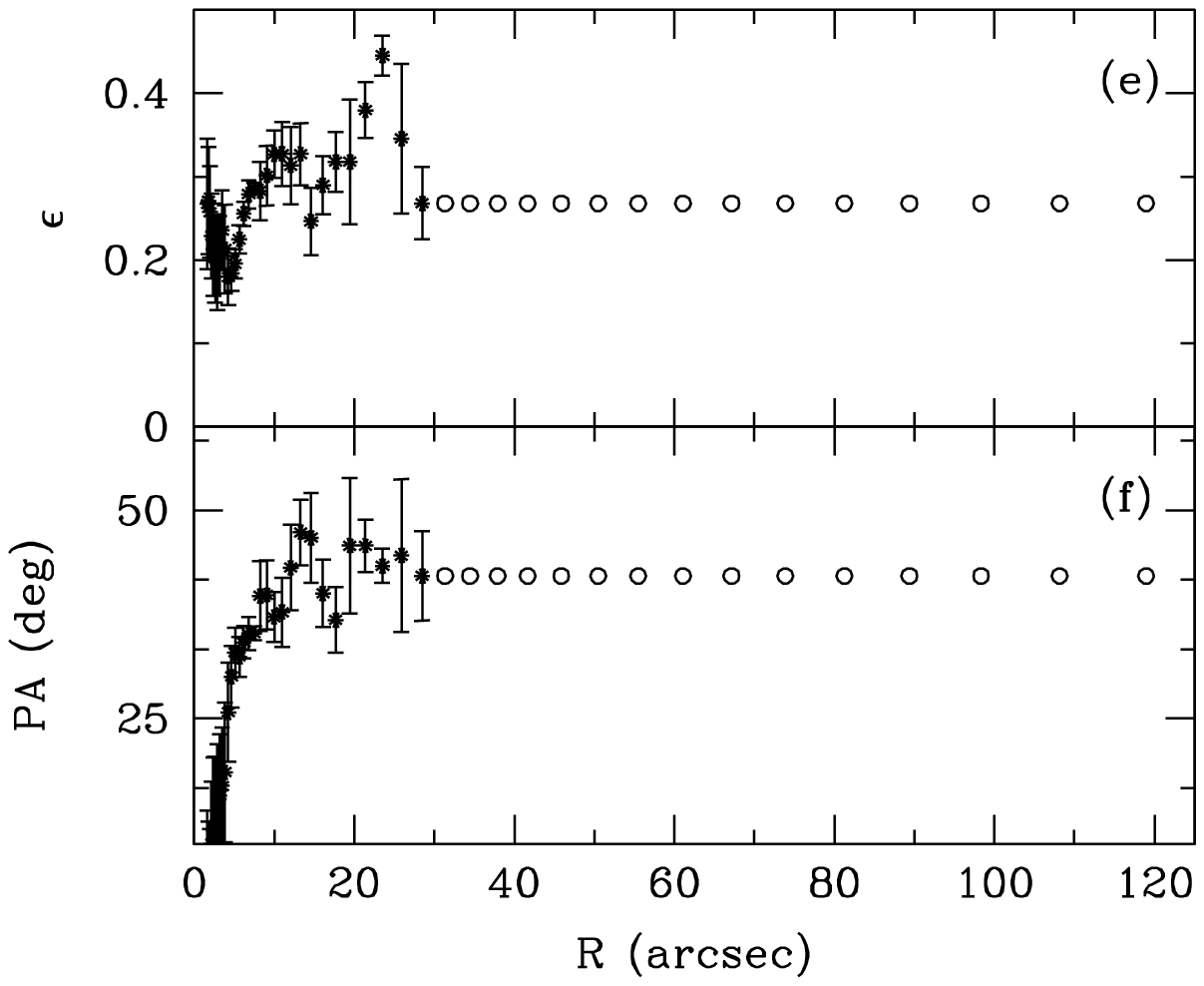}\\
\vspace*{3mm}
\includegraphics[width=6.5cm]{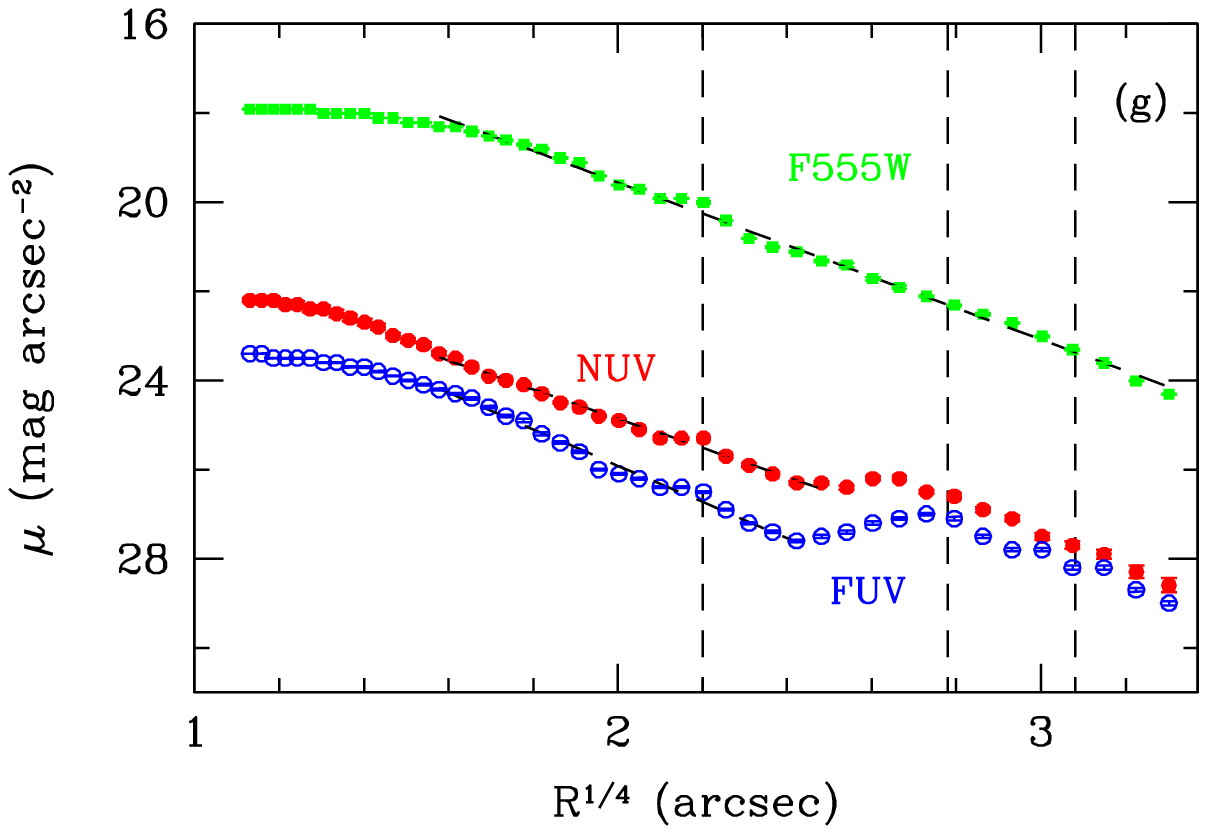}\\
\caption{UV and optical surface photometry NGC~2974, from ellipse
  fitting. {\em (a)} F555W, NUV and FUV surface brightness
  profiles. The horizontal line shows the nominal UV surface
  brightness limit.  {\em (b)--(d)} FUV$-$NUV, FUV$-$F555W and
  NUV$-$F555W colour profiles. {\em (e)--(f)} Ellipticity and position
  angle profiles derived from the NUV image. {\em (g)} F555W, NUV and
  FUV surface brightness profiles on a logarithmic radius scale. The
  features possibly associated with a large-scale bar
  ($R\approx25\arcsec$), the outer ring ($R\approx60\arcsec$) and the
  partial ring ($R\approx90\arcsec$) are marked by vertical
  lines. $R^{1/4}$ fits are shown as straight lines (see text). Open
  circles in (e)--(f) indicate that the values were fixed, not fitted
  or measured.}
\label{fig:rprof}
\end{center}
\end{figure}

Figure~\ref{fig:rprof} shows the multi-colour surface photometry of
NGC~2974 (all magnitudes are AB magnitudes; see also
Table~\ref{tab:rprof}). In Figure~\ref{fig:rprof}a, the dashed line at
m$_{\rm AB}=28$~mag~arcsec$^{-2}$ shows the nominal UV surface
brightness limit for an exposure of $1477$~s. The F555W and UV surface
brightness profiles (Fig.~\ref{fig:rprof}a) show a relatively smooth
decline with radius but also a small peak or plateau between $20$ and
$25\arcsec$. This peak may be associated with the large-scale bar
postulated by \citet{kcemz05} and in
Section~\ref{sec:fig_rot}. Indeed, \citet{wp91} found that the
luminosity profiles of early-type barred galaxies all showed similar
characteristic bumps, correlated with ellipticity peaks. Barred
galaxies are also known to exhibit surface brightness plateaus
\citep*[Freeman Type~II profiles; e.g.][]{f70,mch03,baadbf06}. More
strikingly, a broad secondary peak is clearly observed at
$R\approx60\arcsec$, particularly in the FUV, and can be identified
with the outer ring. The partial ring at $R\approx90\arcsec$ is also
visible as a minor UV peak, but the S/N is very low.
Figure~\ref{fig:rprof}g shows that, well inside the outer ring, the
profile is roughly consistent with a de Vaucouleurs $R^{1/4}$ law
\citep{v58} in all bands. This is also the case at larger radii in the
optical. The optical fit in Figure~\ref{fig:rprof}g is based on all
data points with $R>6\arcsec$, while the FUV and NUV fits are based on
data with $6\arcsec<R<40\arcsec$ and $6\arcsec<R<35\arcsec$,
respectively. The lower bound on the radius is necessary to avoid the
bulk of the seeing effects.
%
%

%
%

\begin{figure}
\begin{center}
\includegraphics[width=8.2cm]{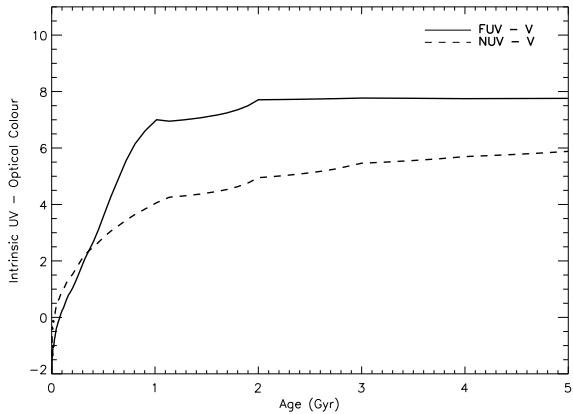}
\caption{UV$-$optical colours of single stellar populations as a
  function of age \citep[see][]{yi03}. Solid and dashed lines show,
  respectively, the FUV$-V$ and NUV$-V$ colour. The evolution in
  UV$-$optical colours slows down dramatically after $1$~Gyr.}
  \label{fig:intcolour}
\end{center}
\end{figure}

\begin{table*}
\caption{Radial profiles based on the ellipse fits}
\label{tab:rprof}
\begin{tabular}{@{}rrrrrrrr}
\hline
Radius & $\mu_{\rm FUV}$ & $\mu_{\rm NUV}$ & $\mu_{\rm F555W}$ & $\log(t_{\rm YC})$ & $\log(f_{\rm YC})$ & $\log(\Sigma_{\rm YC})$ & $\log(\Sigma_{\rm VYC})$ \\
(\arcsec) & (mag~arcsec$^{-2}$) & (mag~arcsec$^{-2}$) & (mag~arcsec$^{-2}$) & (yr) &  & (M$_\odot$~arcsec$^{-2}$) & (M$_\odot$~arcsec$^{-2}$) \\
\hline
$1.64$ & $23.39\pm0.10$ & $22.19\pm0.09$ & $17.91\pm0.03$ & $8.56^{+0.10}_{-0.10}$ & $-2.10^{+0.40}_{-0.30}$ & $6.15^{+0.06}_{-0.07}$ & $6.15^{+0.06}_{-0.07}$ \\
$1.80$ & $23.43\pm0.11$ & $22.21\pm0.11$ & $17.92\pm0.05$ & $8.81^{+0.00}_{-0.05}$ & $-2.10^{+0.18}_{-0.30}$ & $6.12^{+0.06}_{-0.08}$ & $6.12^{+0.06}_{-0.08}$ \\
$1.98$ & $23.51\pm0.11$ & $22.26\pm0.11$ & $17.92\pm0.05$ & $8.61^{+0.00}_{-0.15}$ & $-2.00^{+0.00}_{-0.40}$ & $6.09^{+0.07}_{-0.08}$ & $6.09^{+0.07}_{-0.08}$ \\
$2.18$ & $23.52\pm0.11$ & $22.32\pm0.12$ & $17.93\pm0.05$ & $8.61^{+0.00}_{-0.15}$ & $-2.00^{+0.00}_{-0.40}$ & $6.05^{+0.07}_{-0.08}$ & $6.05^{+0.07}_{-0.08}$ \\
$2.39$ & $23.54\pm0.11$ & $22.35\pm0.12$ & $17.95\pm0.05$ & $8.61^{+0.00}_{-0.15}$ & $-2.00^{+0.00}_{-0.40}$ & $5.99^{+0.07}_{-0.09}$ & $5.99^{+0.07}_{-0.09}$ \\
$2.63$ & $23.57\pm0.11$ & $22.41\pm0.12$ & $17.97\pm0.05$ & $8.61^{+0.00}_{-1.20}$ & $-2.05^{+0.05}_{-0.48}$ & $5.93^{+0.08}_{-0.09}$ & $5.93^{+0.08}_{-0.09}$ \\
$2.90$ & $23.64\pm0.11$ & $22.47\pm0.13$ & $18.01\pm0.07$ & $8.61^{+0.00}_{-0.15}$ & $-2.05^{+0.05}_{-0.48}$ & $5.87^{+0.08}_{-0.10}$ & $5.87^{+0.08}_{-0.10}$ \\
$3.18$ & $23.69\pm0.11$ & $22.52\pm0.13$ & $18.03\pm0.07$ & $8.61^{+0.05}_{-0.15}$ & $-2.10^{+0.10}_{-0.43}$ & $5.81^{+0.09}_{-0.12}$ & $5.81^{+0.09}_{-0.12}$ \\
$3.50$ & $23.72\pm0.11$ & $22.66\pm0.13$ & $18.05\pm0.07$ & $8.51^{+0.15}_{-0.30}$ & $-2.40^{+0.40}_{-0.60}$ & $5.74^{+0.11}_{-0.14}$ & $5.74^{+0.11}_{-0.14}$ \\
$3.86$ & $23.77\pm0.11$ & $22.72\pm0.12$ & $18.07\pm0.07$ & $8.51^{+0.15}_{-0.30}$ & $-2.40^{+0.40}_{-0.60}$ & $5.63^{+0.12}_{-0.16}$ & $5.63^{+0.12}_{-0.16}$ \\
$4.23$ & $23.84\pm0.11$ & $22.87\pm0.11$ & $18.12\pm0.07$ & $8.41^{+0.20}_{-0.35}$ & $-2.70^{+0.60}_{-0.60}$ & $5.48^{+0.11}_{-0.15}$ & $5.48^{+0.11}_{-0.15}$ \\
$4.65$ & $23.92\pm0.11$ & $23.01\pm0.11$ & $18.16\pm0.07$ & $8.51^{+0.10}_{-1.00}$ & $-2.42^{+0.30}_{-1.48}$ & $5.41^{+0.11}_{-0.15}$ & $5.41^{+0.11}_{-0.15}$ \\
$5.13$ & $24.08\pm0.11$ & $23.15\pm0.10$ & $18.21\pm0.07$ & $8.46^{+0.15}_{-0.90}$ & $-2.70^{+0.40}_{-1.30}$ & $5.13^{+0.12}_{-0.17}$ & $5.13^{+0.12}_{-0.17}$ \\
$5.64$ & $24.14\pm0.12$ & $23.24\pm0.10$ & $18.25\pm0.07$ & $8.46^{+0.15}_{-0.90}$ & $-2.70^{+0.40}_{-1.30}$ & $5.21^{+0.11}_{-0.15}$ & $5.21^{+0.11}_{-0.15}$ \\
$6.20$ & $24.23\pm0.12$ & $23.46\pm0.10$ & $18.31\pm0.07$ & $7.70^{+0.86}_{-0.11}$ & $-4.00^{+1.48}_{-0.00}$ & $5.05^{+0.11}_{-0.14}$ & $5.05^{+0.11}_{-0.14}$ \\
$6.83$ & $24.36\pm0.13$ & $23.58\pm0.09$ & $18.37\pm0.07$ & $7.72^{+0.79}_{-0.09}$ & $-4.00^{+1.30}_{-0.00}$ & $4.89^{+0.11}_{-0.16}$ & $4.89^{+0.11}_{-0.16}$ \\
$7.50$ & $24.47\pm0.13$ & $23.73\pm0.09$ & $18.43\pm0.07$ & $7.76^{+0.65}_{-0.08}$ & $-4.00^{+1.00}_{-0.00}$ & $4.77^{+0.12}_{-0.17}$ & $4.77^{+0.12}_{-0.17}$ \\
$8.25$ & $24.62\pm0.13$ & $23.91\pm0.08$ & $18.55\pm0.07$ & $7.81^{+0.60}_{-0.11}$ & $-4.00^{+1.00}_{-0.00}$ & $4.41^{+0.09}_{-0.12}$ & $4.41^{+0.09}_{-0.12}$ \\
$9.08$ & $24.84\pm0.13$ & $24.04\pm0.08$ & $18.62\pm0.08$ & $7.86^{+0.60}_{-0.12}$ & $-4.00^{+1.00}_{-0.00}$ & $4.36^{+0.07}_{-0.09}$ & $4.36^{+0.07}_{-0.09}$ \\
$9.99$ & $24.95\pm0.13$ & $24.13\pm0.08$ & $18.71\pm0.08$ & $7.86^{+0.60}_{-0.12}$ & $-4.00^{+1.00}_{-0.00}$ & $4.65^{+0.08}_{-0.15}$ & $4.65^{+0.08}_{-0.15}$ \\
$10.98$ & $25.25\pm0.13$ & $24.35\pm0.08$ & $18.83\pm0.08$ & $7.96^{+0.55}_{-0.01}$ & $-4.00^{+1.00}_{-0.00}$ & $4.54^{+0.06}_{-0.07}$ & $4.54^{+0.06}_{-0.07}$ \\
$12.08$ & $25.46\pm0.13$ & $24.52\pm0.08$ & $19.02\pm0.08$ & $7.96^{+0.55}_{-0.14}$ & $-4.00^{+1.00}_{-0.00}$ & $4.77^{+0.07}_{-0.08}$ & $4.77^{+0.06}_{-0.07}$ \\
$13.29$ & $25.67\pm0.13$ & $24.67\pm0.08$ & $19.17\pm0.08$ & $7.96^{+0.55}_{-0.01}$ & $-4.00^{+1.00}_{-0.00}$ & $4.84^{+0.06}_{-0.06}$ & $4.77^{+0.06}_{-0.06}$ \\
$14.61$ & $26.04\pm0.12$ & $24.85\pm0.08$ & $19.42\pm0.06$ & $8.41^{+0.25}_{-0.55}$ & $-3.30^{+0.78}_{-0.70}$ & $4.79^{+0.05}_{-0.05}$ & $4.61^{+0.05}_{-0.05}$ \\
$16.08$ & $26.13\pm0.12$ & $24.99\pm0.08$ & $19.65\pm0.06$ & $8.36^{+0.25}_{-0.55}$ & $-3.30^{+0.78}_{-0.70}$ & $4.86^{+0.05}_{-0.06}$ & $4.58^{+0.05}_{-0.06}$ \\
$17.69$ & $26.25\pm0.12$ & $25.12\pm0.09$ & $19.78\pm0.06$ & $8.36^{+0.25}_{-0.55}$ & $-3.30^{+0.78}_{-0.70}$ & $4.80^{+0.05}_{-0.06}$ & $4.57^{+0.05}_{-0.06}$ \\
$19.46$ & $26.42\pm0.12$ & $25.32\pm0.09$ & $19.91\pm0.06$ & $8.36^{+0.25}_{-0.55}$ & $-3.30^{+0.60}_{-0.70}$ & $4.75^{+0.05}_{-0.05}$ & $4.47^{+0.05}_{-0.05}$ \\
$21.41$ & $26.49\pm0.12$ & $25.33\pm0.08$ & $19.94\pm0.06$ & $8.36^{+0.30}_{-0.50}$ & $-3.52^{+1.00}_{-0.48}$ & $4.85^{+0.04}_{-0.04}$ & $4.61^{+0.04}_{-0.04}$ \\
$23.54$ & $26.53\pm0.12$ & $25.35\pm0.09$ & $20.02\pm0.09$ & $8.36^{+0.30}_{-0.55}$ & $-3.30^{+0.78}_{-0.70}$ & $4.77^{+0.04}_{-0.05}$ & $4.56^{+0.04}_{-0.05}$ \\
$25.89$ & $26.91\pm0.12$ & $25.71\pm0.09$ & $20.45\pm0.07$ & $8.46^{+0.20}_{-0.65}$ & $-3.00^{+0.60}_{-1.00}$ & $4.85^{+0.04}_{-0.05}$ & $4.27^{+0.04}_{-0.05}$ \\
$28.49$ & $27.25\pm0.13$ & $25.93\pm0.09$ & $20.88\pm0.07$ & $8.51^{+0.20}_{-0.30}$ & $-2.70^{+0.60}_{-0.70}$ & $4.84^{+0.04}_{-0.04}$ & $3.77^{+0.04}_{-0.04}$ \\
$31.34$ & $27.46\pm0.13$ & $26.17\pm0.09$ & $21.06\pm0.08$ & $8.61^{+0.05}_{-0.35}$ & $-2.52^{+0.22}_{-0.78}$ & $4.88^{+0.04}_{-0.05}$ & $3.87^{+0.04}_{-0.04}$ \\
$34.46$ & $27.63\pm0.13$ & $26.32\pm0.10$ & $21.14\pm0.10$ & $8.56^{+0.10}_{-0.70}$ & $-2.70^{+0.40}_{-1.30}$ & $4.90^{+0.04}_{-0.04}$ & $3.48^{+0.04}_{-0.04}$ \\
$37.91$ & $27.54\pm0.13$ & $26.36\pm0.09$ & $21.38\pm0.09$ & $8.56^{+0.10}_{-0.40}$ & $-2.52^{+0.37}_{-0.78}$ & $4.70^{+0.04}_{-0.04}$ & $3.84^{+0.04}_{-0.04}$ \\
$41.70$ & $27.41\pm0.14$ & $26.41\pm0.10$ & $21.43\pm0.10$ & $8.46^{+0.15}_{-0.87}$ & $-2.70^{+0.40}_{-1.30}$ & $4.72^{+0.04}_{-0.05}$ & $3.86^{+0.04}_{-0.05}$ \\
$45.87$ & $27.23\pm0.14$ & $26.23\pm0.10$ & $21.79\pm0.10$ & $8.56^{+0.05}_{-0.40}$ & $-2.15^{+0.15}_{-0.85}$ & $4.65^{+0.04}_{-0.04}$ & $4.09^{+0.04}_{-0.04}$ \\
$50.46$ & $27.18\pm0.13$ & $26.24\pm0.10$ & $21.94\pm0.09$ & $8.56^{+0.05}_{-0.30}$ & $-2.05^{+0.05}_{-0.65}$ & $4.73^{+0.04}_{-0.04}$ & $4.26^{+0.04}_{-0.04}$ \\
$55.50$ & $27.02\pm0.13$ & $26.53\pm0.09$ & $22.12\pm0.08$ & $7.86^{+0.65}_{-0.64}$ & $-3.30^{+1.15}_{-0.70}$ & $4.50^{+0.03}_{-0.04}$ & $4.30^{+0.03}_{-0.04}$ \\
$61.05$ & $27.12\pm0.14$ & $26.67\pm0.10$ & $22.35\pm0.08$ & $8.06^{+0.45}_{-0.88}$ & $-3.00^{+0.90}_{-1.00}$ & $4.47^{+0.04}_{-0.04}$ & $4.14^{+0.04}_{-0.04}$ \\
$67.16$ & $27.53\pm0.14$ & $26.92\pm0.11$ & $22.52\pm0.08$ & $8.26^{+0.30}_{-1.01}$ & $-2.70^{+0.65}_{-1.30}$ & $4.56^{+0.04}_{-0.04}$ & $3.92^{+0.04}_{-0.04}$ \\
$73.88$ & $27.89\pm0.14$ & $27.14\pm0.12$ & $22.78\pm0.08$ & $8.46^{+0.10}_{-0.60}$ & $-2.30^{+0.30}_{-1.00}$ & $4.50^{+0.05}_{-0.05}$ & $3.29^{+0.05}_{-0.05}$ \\
$81.26$ & $27.85\pm0.14$ & $27.56\pm0.13$ & $23.05\pm0.08$ & $7.30^{+1.06}_{-0.08}$ & $-4.00^{+1.48}_{-0.00}$ & $3.82^{+0.06}_{-0.07}$ & $3.37^{+0.06}_{-0.07}$ \\
$89.39$ & $28.23\pm0.15$ & $27.78\pm0.14$ & $23.32\pm0.08$ & $7.86^{+0.60}_{-0.62}$ & $-3.30^{+1.00}_{-0.70}$ & $3.41^{+0.09}_{-0.11}$ & $2.60^{+0.09}_{-0.11}$ \\
$98.33$ & $28.26\pm0.14$ & $27.91\pm0.15$ & $23.65\pm0.06$ & $8.00^{+0.45}_{-0.87}$ & $-3.00^{+0.78}_{-1.00}$ & $2.41^{+0.13}_{-0.19}$ & $1.83^{+0.13}_{-0.19}$ \\
$108.15$ & $28.71\pm0.14$ & $28.33\pm0.19$ & $24.04\pm0.06$ & $7.81^{+0.65}_{-065}$ & $-3.30^{+1.08}_{-0.70}$ & $           -       $ & $            -       $ \\
$118.98$ & $29.02\pm0.15$ & $28.61\pm0.22$ & $24.32\pm0.06$ & $7.81^{+0.65}_{-0.65}$ & $-3.30^{+1.08}_{-0.70}$ & $           -       $ & $            -       $ \\
\hline
\end{tabular}
Columns: (1) Major-axis radius; (2)--(4) surface brightnesses in the
FUV, NUV and F555W filters; (5)--(6) age and mass fraction of the
young component; (7) surface mass densities of the young component
from the pixel-by-pixel analysis shown in Figure~\ref{fig:mass}; (8)
same as (7) but including only the very young stars (age $<0.5$~Gyr).
\end{table*}
Figure~\ref{fig:intcolour} shows that a population of age $\la0.5$~Gyr
would exhibit FUV$-$NUV$\la1.0$, which is seen only in the central
region ($R\la10\arcsec$) and near and beyond the newly-detected outer
ring ($R\ga45\arcsec$). Studies of large samples of elliptical
galaxies have indeed suggested that some have blue colours
\citep[e.g.][]{bbb93} which can be explained by new generations of
stars, and young stars are effectively traced by UV$-$optical
colours. Using the empirical criteria of \citet{yetal05} (based on
NGC~4552 and M~32), thresholds below which we can reasonably expect
the colours to indicate young stars (age$\,\la1$~Gyr) are shown as
dashed lines in Figure~\ref{fig:rprof}b--d: FUV$-$NUV$=1.0$,
FUV$-$F555W$=6.1$ and NUV$-$F555W$=5.1$. Considering those
UV$-$optical colours, it seems that young stellar populations are
present in the core and around the outer ring, implying recent star
formation in those two locations, despite the lack of other star
formation evidence and the standard classification of NGC~2974 as an
E4 elliptical. A note of caution is however necessary as strong FUV
emission in the central parts does not always indicate RSF. The
FUV$-$NUV and FUV$-$F555W colours in these regions are close to those
of typical UV upturn galaxies (e.g.\ NGC~1399, \citealt[][]{betal96};
NGC~4552, \citealt[][]{fetal91}), where old helium-burning stars are
believed to be the primary FUV sources. The UV bright outer ring is
however difficult to explain with He-burning stars, as the UV upturn
phenomenon is generally found in galaxy cores.
%
%

The morphology of NGC~2974 is also revealing
(Fig.~\ref{fig:rprof}e--h). The NUV ellipticity is somewhat noisy
outside of the central region ($R\ga10\arcsec$) but is consistent with
a single value $\epsilon\approx0.35$. The ellipticity drops
continuously at smaller radii, reaching $\epsilon\approx0.15$ in the
centre. This suggests that the two components seen in the FUV and NUV
images have different flattening, the inner component being rounder
(i.e.\ spheroid-like) and the outer component flatter (i.e.\
disc-like). Some evidence in fact suggest that NGC~2974 has a stellar
disc, although it is probably quite thick
\citep[e.g.][]{kcemz05}. While invalid for this thick disc and the
stellar spheroid traced, e.g., by the optical light, the thin disc
approximation is probably acceptable for the UV light tracing (very)
young stars (see Section~\ref{sec:spop_sf}), at the very least in the
outer ring, since those stars were presumably born in a thin gaseous
disc. For an infinitely thin disc, $\epsilon=0.35$ implies an
inclination $i=49\degr$, somewhat smaller than the values of
$55$--$65\degr$ derived by previous studies
\citep[e.g.][]{kjgkg88,aetal93,betal93,cm94,pbacm98,kcemz05}.
However, ellipse fitting fails at large radii. Measuring the axial
ratio of the outer ring directly on the FUV image yields an
ellipticity $\epsilon=0.43$, corresponding to an inclination of
$55\degr$, in better agreement with previous works. It is thus likely
that the newly-identified UV rings lie in a relatively thin disc in
the equatorial plane of NGC~2974.

The NUV position angle is also noisy but roughly constant outside the
central region. At $R\la10\arcsec$, however, the position angle varies
continuously. This isophotal twist suggests an asymmetry in the
central region, consistent with the presence of a nuclear bar or
nuclear spiral arms, although it could equally be due to dust
\citep[e.g.][]{egf03}. The NUV Fourier coefficients $a_3$ and $a_4$,
which measure deviations of the isophotes from a pure ellipse
\citep[e.g.][]{l85,bm87,j87}, are essentially consistent with zero at
all radii (not shown).
%
%

\begin{figure}
\begin{center}
\includegraphics[width=8cm]{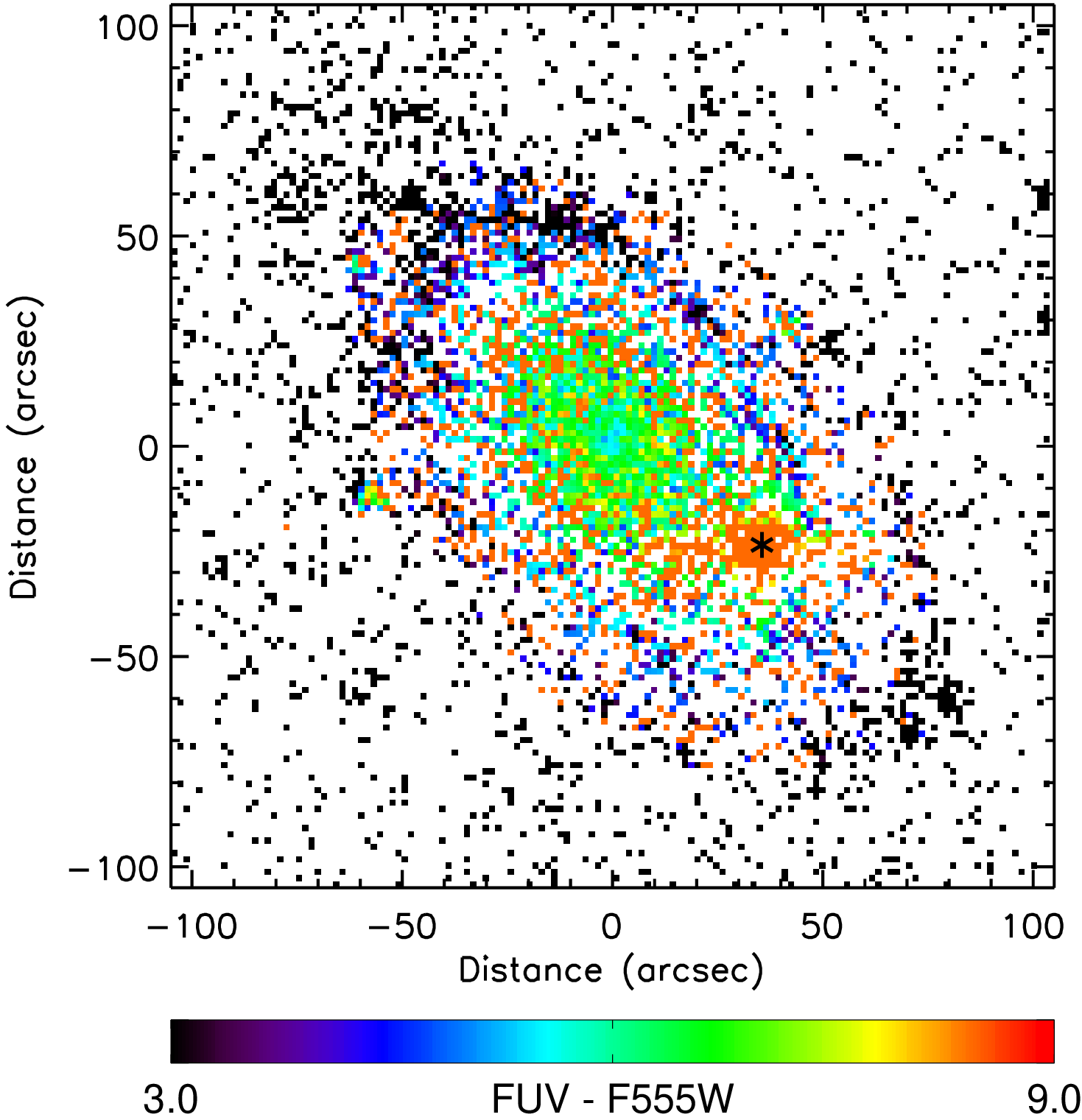}\\
\vspace*{3mm}
\includegraphics[width=8cm]{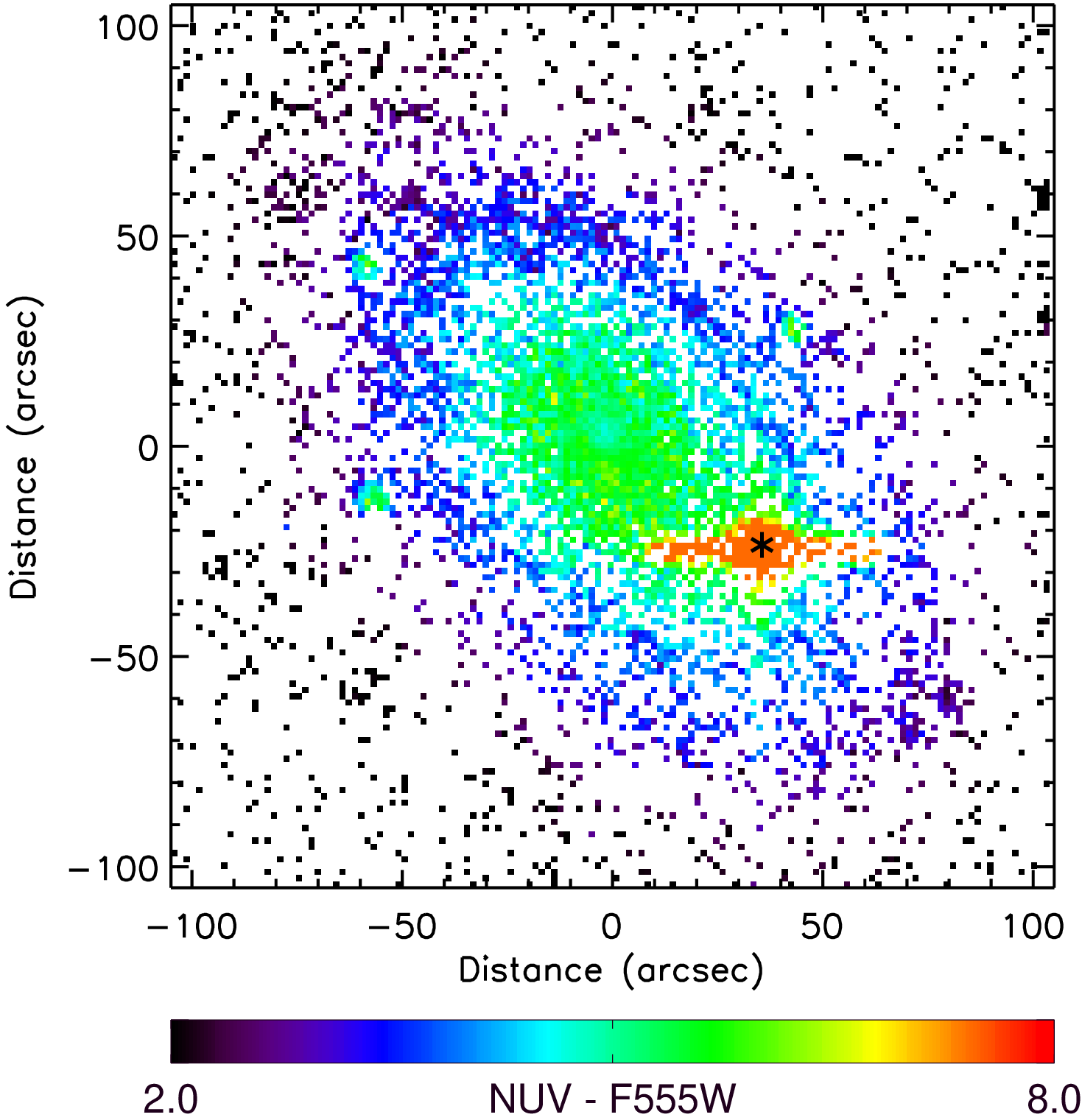}
\caption{UV$-$optical colour maps of NGC~2974. {\em Top:} FUV$-$F555W.
  {\em Bottom:} NUV$-$F555W. A star marks the area contaminated by the
  foreground star.}
\label{fig:colourmaps}
\end{center}
\end{figure}
%
%

\begin{figure*}
\begin{center}
\includegraphics[width=17cm]{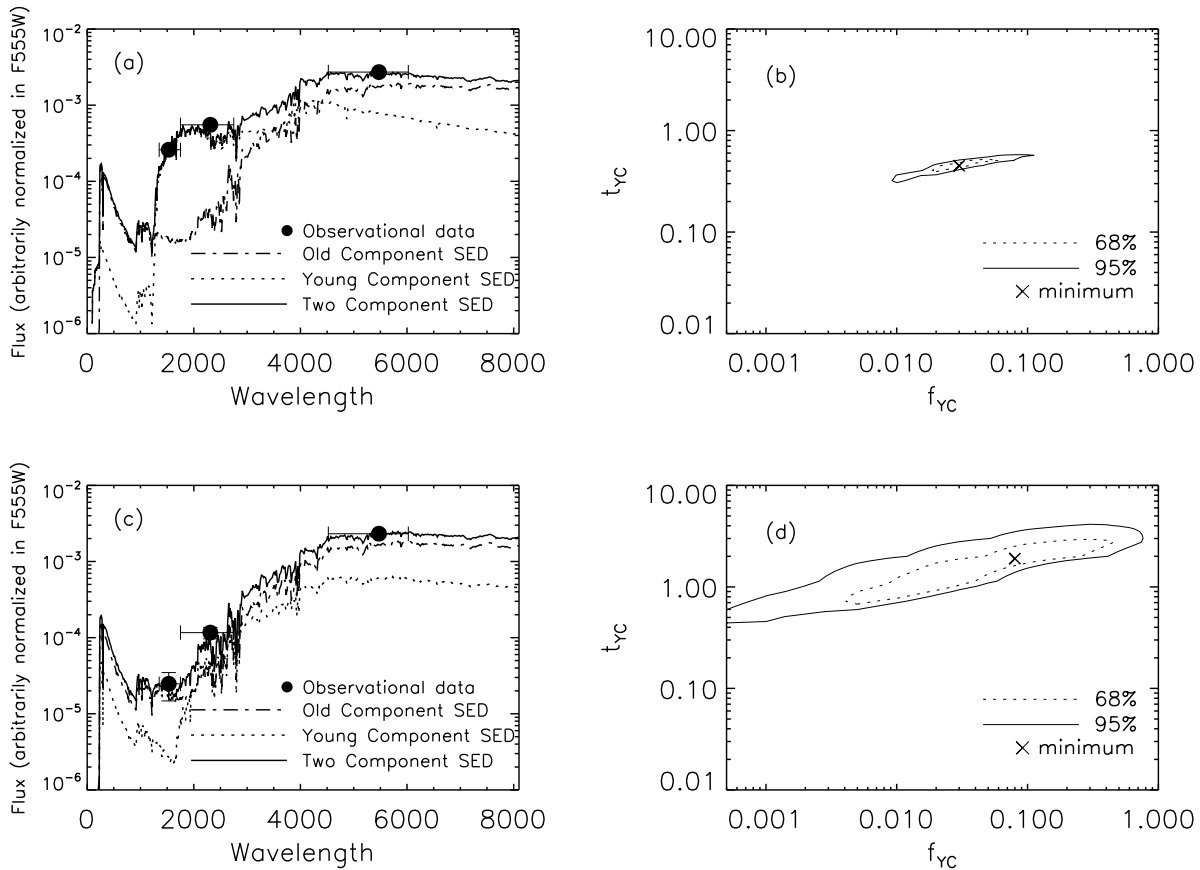}
\caption{Examples of two-component stellar population fits. {\em (a)}
Model spectra, where the dotted and dot-dashed spectra represent the
young and old component, respectively, while the solid spectrum is the
sum. {\em (b)} $\chi^2$ contours, where the best fit is marked with an
`x'. The dotted and dot-dashed contours represent the $68$ and
$95$~per cent confidence levels, respectively. This example from the
outer ring has a non-negligible fraction ($\approx3$~per cent) of very
young ($0.45$~Gyr) stars. {\em (c)}--{\em (d)} Same as (a)--(b) but
for an example at intermediate radius ($R\approx35\arcsec$), where the
age and mass fraction of the young component are poorly constrained.}
\label{fig:fit_ex}
\end{center}
\end{figure*}
%
%
\section{STELLAR POPULATIONS AND STAR FORMATION}
\label{sec:spop_sf}
\subsection{Distribution of young and old stars}
\label{sec:young_old}
\citet{ba44} proposed the existence of two kinds of stellar
populations by analyzing the colours and brightnesses of nearby spiral
and elliptical galaxies: population~I, containing highly luminous O
and B type stars; and population~II, containing luminous red stars. He
also suggested that elliptical galaxies consist of population II stars
in a gas and dust-free environment. In the past sixty years, however,
many surveys have detected (extended) gaseous discs in ellipticals,
and atomic \citep[e.g.][]{kjgkg88,wkomz06}, ionised
\citep*[e.g.][]{b92,dbb84,k89,ghjn94,pbacm98} and X-ray
\citep*[e.g.][]{fjt85} gas have all been detected in NGC~2974. Detailed
studies of the colours and brightnesses of stars revealed that the
distinction between populations~I and II is also one of age. Some
elliptical galaxies (including NGC~2974) have blue UV colours, and
this is believed to be caused mainly by young stars (see
Section~\ref{sec:results}).

Most UV photons are emitted by stars younger than $\sim10^9$~yr, so
the UV luminosity is closely related to the RSF history of the
galaxy. While radial profiles such as those shown in
Figure~\ref{fig:rprof} are very useful, they do not allow to study the
detailed spatial distribution of young stars, and so we have
constructed full two-dimensional UV$-$optical colour maps, shown in
Figure~\ref{fig:colourmaps}. The FUV$-$NUV colour map is very noisy,
and thus in the following analysis we restrict ourselves to pixels
with S/N$>2$ in both FUV and NUV.

The FUV$-$F555W colour map clearly shows the blue colour
(FUV$-$F555W$\la6.1$) of the outer ring and outskirts of NGC~2974,
suggesting that young stars are an important contributor to the
galaxy's UV luminosity. With FUV$-$F555W$\approx5.5$, the central
regions also likely harbour young stars, while the intermediate
regions are red (FUV$-$F555W$\ga6.1$). Those trends are fully
supported by the NUV$-$F555W colours, where NUV$-$F555W$\la5.1$ traces
young stars. It also appears that there are more young stars in the
North-East half of the outer ring than to the South-West.
\subsection{Age and mass fraction of young stars}
\label{sec:age_mass}
In order to estimate the age, mass fraction and surface mass density
of the young stellar component, we consider a two-stage star formation
history, whereby stars are formed instantaneously at two different
times. The first starburst is fixed at high redshift ($12$~Gyr) and
represents an old stellar component. It is based on a
metallicity-composite population (a short burst with chemical
enrichment) with a mean metallicity of roughly solar, which includes
the UV upturn phenomenon originating from old hot HB stars
\citep[see][]{ka97}. The second burst has a fixed metallicity (solar)
but its age and relative magnitude are left to vary. The free
parameters of the model are thus the age ($t_{\rm YC}$) and stellar
mass fraction ($f_{\rm YC}$) of the young stellar component, and we
explore a wide range of values for both: $10^{-3}\le t_{\rm
YC}\le10$~Gyr and $10^{-6}\le f_{\rm YC}\le1$.  The surface mass
density of the young component ($\Sigma_{\rm YC}$) is then calculated
by simply multiplying $f_{\rm YC}$ by the total surface mass density,
obtained here directly from the F555W surface brightness assuming a
purely old population, a reasonable approximation given the small
values of $f_{\rm YC}$ derived everywhere.

The models of \citet{yi03}, updated to allow a wider choice of initial
mass functions (Yi \& Yoon, in prep.), are used. Since those models do
not cover ages younger than $1$~Gyr, we combine them with the models
of \citet{bc03} at an age of $1$~Gyr. The connection between the two
sets of models is reasonably smooth, as illustrated in
Figure~\ref{fig:intcolour} which shows the intrinsic UV$-V$ colours of
a simple stellar population (SSP) as a function of age. Clearly, small
values of UV$-$optical colours trace young stars. Furthermore, the UV
emission in star-forming galaxies is dominated by short-lived stars
\citep[e.g.][]{ke98}, so the evolution of the UV$-$optical colours
slows down dramatically after $1$~Gyr.

To determine the age and the mass fraction of the young component, we
fit the observed UV and optical colours (FUV$-$F555W and NUV$-$F555W)
to those returned by the two-component model. Dust extinction is
expected to be small everywhere in NGC~2974, so for simplicity we do
not correct for it. The mass fraction of the young component is then
formally a lower limit, since dust would mainly affect the FUV and NUV
filters. For illustration, Figure~\ref{fig:fit_ex} shows the fit and
$\chi^2$ contours ($68$~per cent and $95$~per cent confidence levels)
for two different locations (pixels) in the galaxy. The dotted and dot-dashed
spectra represent the young and old component, respectively, while the
solid spectrum is the sum. In the first case, taken in the outer ring,
there is strong evidence for a stellar component younger than
$0.5$~Gyr. In the second, taken at intermediate radius
($R=35\arcsec$), the age and mass fraction of the young component are
poorly constrained. The shape of the $\chi^2$ contours illustrates the
usual age-mass degeneracy, whereby a small amount of very young stars
has a similar effect to a larger amount of older stars.
%
%

\begin{figure}
\begin{center}
\includegraphics[width=7cm]{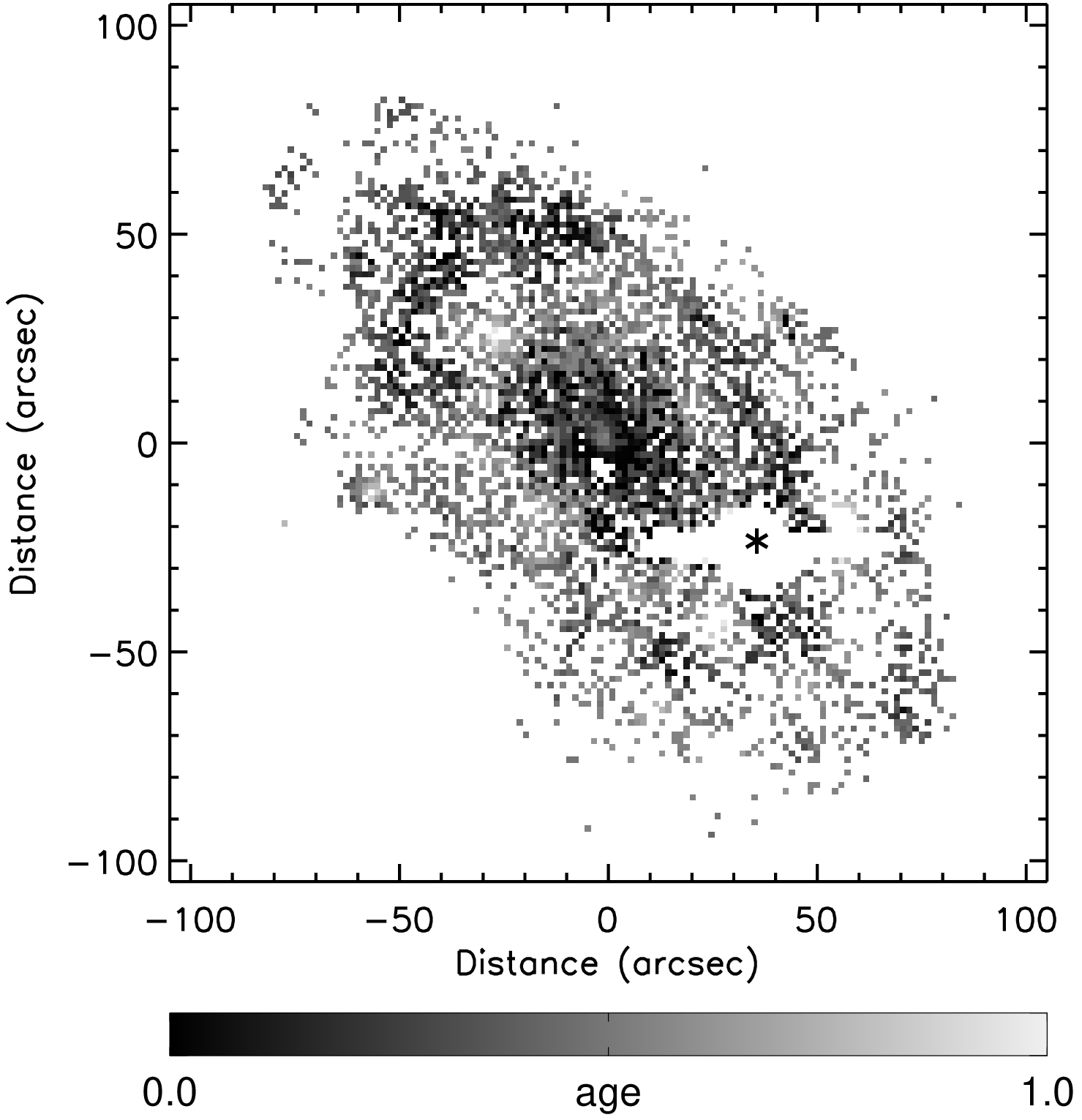}\\
\vspace*{3mm}
\includegraphics[width=7cm]{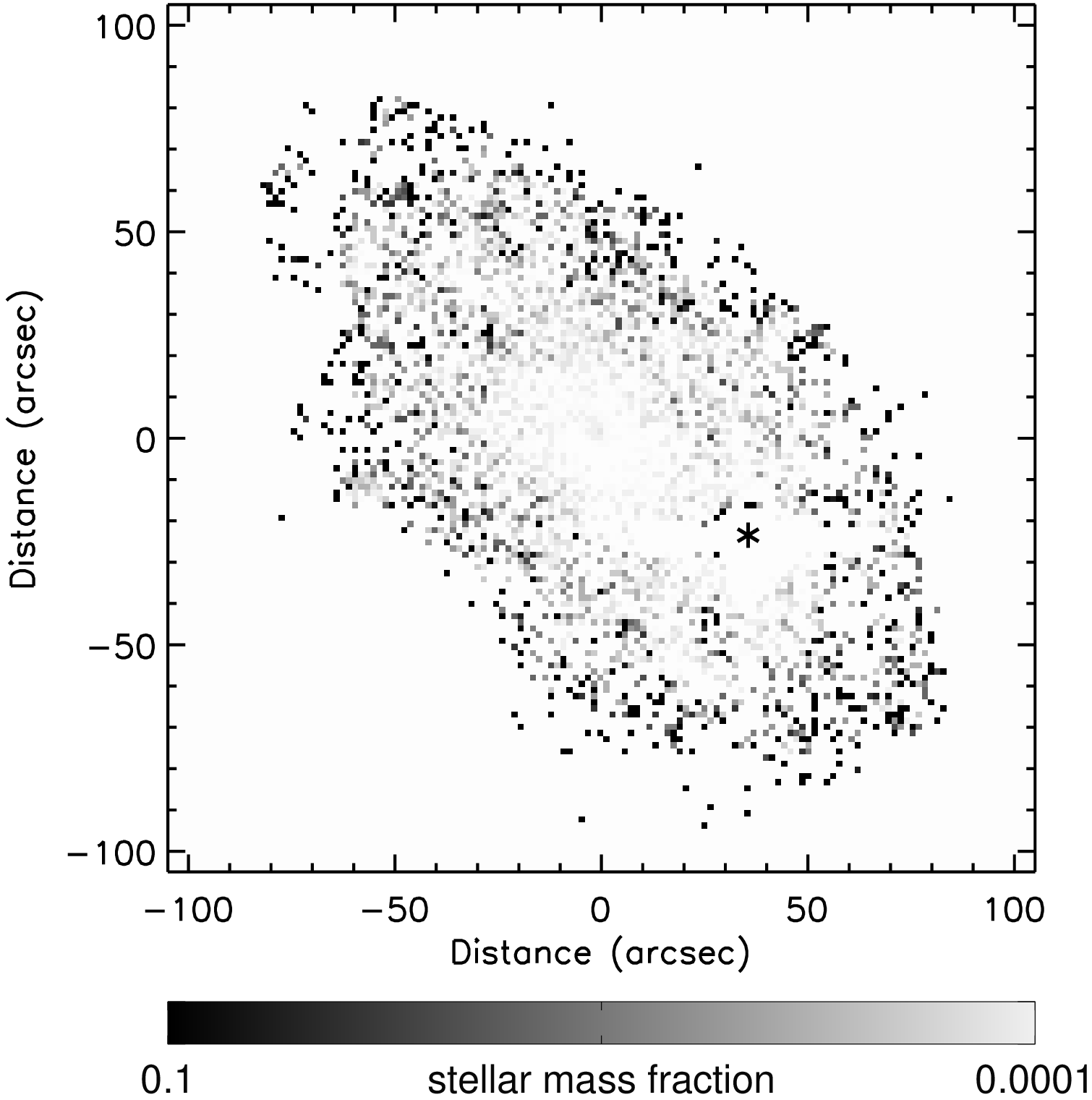}\\
\vspace*{3mm}
\includegraphics[width=7cm]{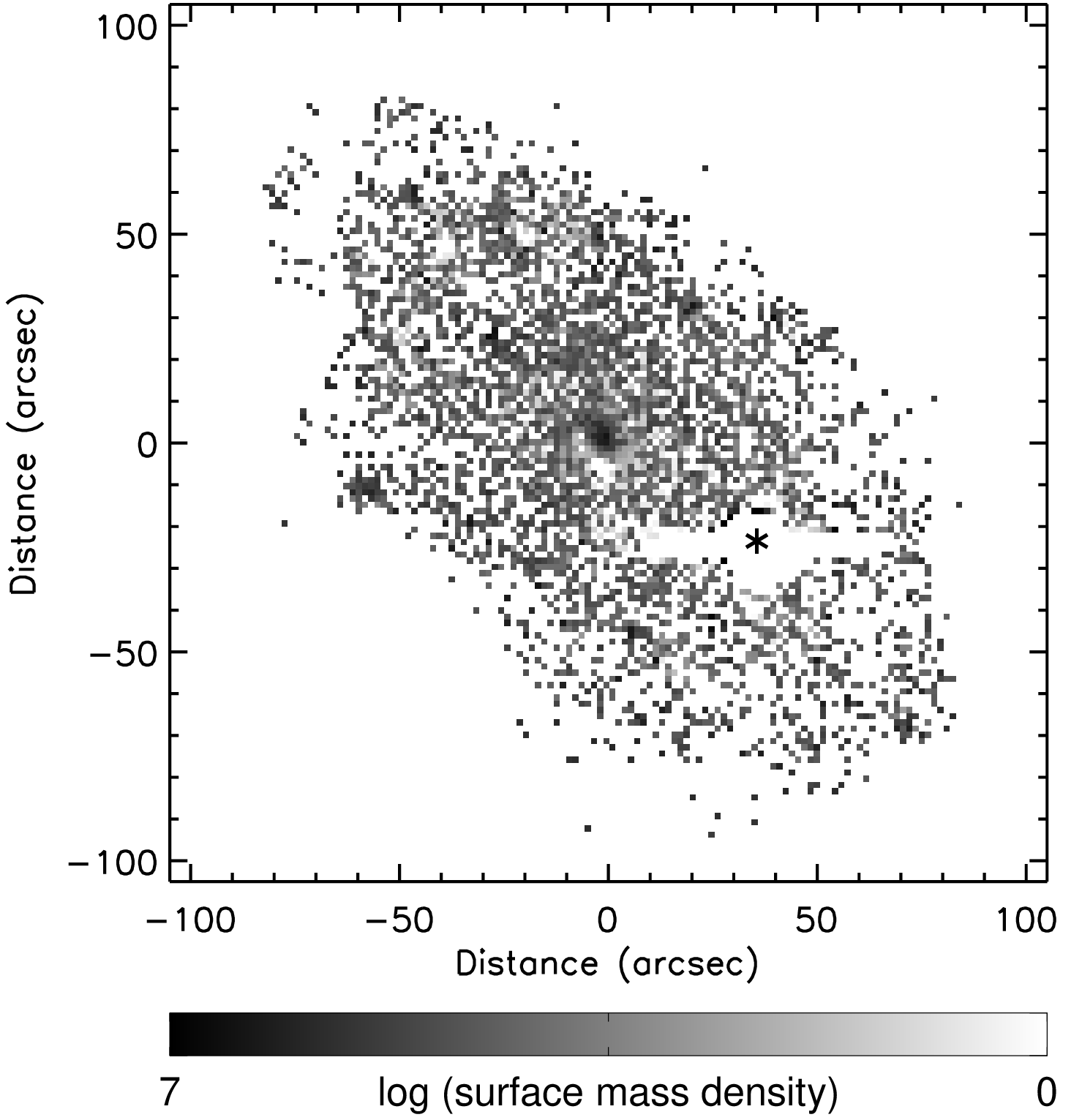}
\caption{Two-component fit maps for the pixels with S/N$>2$. Age {\em
(top)} in unit of Gyr, stellar mass fraction {\em (middle)} and
surface mass density in unit of M$_{\odot}$ arcsec$^{-2}$ {\em
(bottom)} of the young stellar component. A star marks the area
contaminated by the foreground star.} \label{fig:fit_maps}
\end{center}
\end{figure}
In Figure~\ref{fig:fit_maps}, we show the best estimates of ($t_{\rm
YC}$, $f_{\rm YC}$, $\Sigma_{\rm YC}$) using two-dimensional maps.
Pixels with $\chi^2>1.5$ (bad fits) are shown in white, as are fits
with a negligible young component. The age map clearly shows very
young stars ($<500$~Myr) in both the central regions and in and around
the newly-identified UV rings, as suggested by the FUV data (e.g.\
Fig.~\ref{fig:images}). The prominence of the central spheroid
compared to the outer ring in both UV images can probably be explained
by the fact that, while the mass fraction in the young component is
not particularly high in the centre, the total surface mass density
there is approximately $50$ times higher than in the outer ring (e.g.\
Fig.~\ref{fig:rprof}), resulting in a much higher absolute surface
mass density for the young component. Interestingly, there is also a
sprinkling of young stars at intermediate radii, between the inner
spheroid and the outer ring, and there are comparatively more (very)
young stars to the North-East edge of the outer ring than to the
South-West. There is however considerable pixel-to-pixel scatter. The
mass fraction map clearly shows that the fractional contribution of
the young component is rather low in the centre but increases outward,
explaining the very blue colours of NGC~2974 near the outer ring.

\begin{figure}
\begin{center}
\includegraphics[width=8cm]{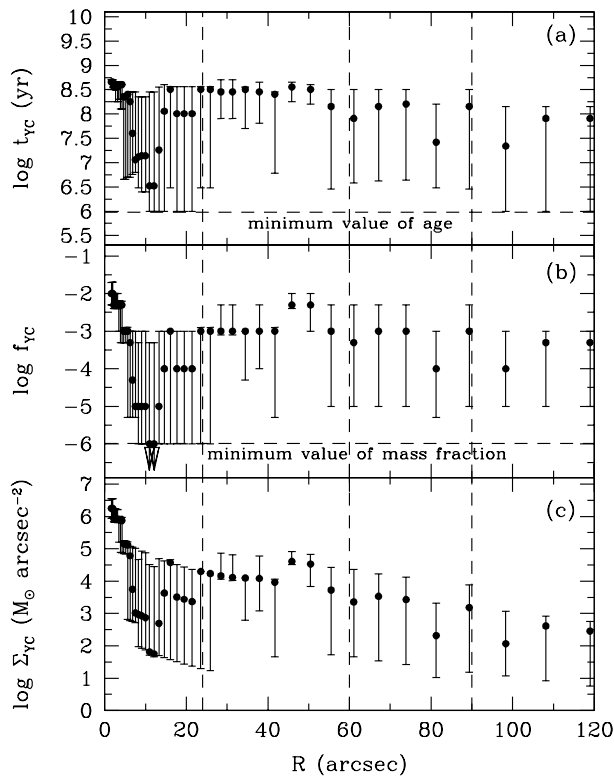}
\caption{Two-component fit radial profiles. Errors show the ranges of
the fitting model parameters that are within 1$\sigma$ confidence from
the best-fit model. Age {\em (a)}, stellar mass fraction {\em (b)} and
surface mass density {\em (c)} of the young stellar component. The
vertical lines show the radial positions of the large-scale bar
($R\approx25\arcsec$), the outer ring ($R\approx60\arcsec$) and the
partial ring ($R\approx90\arcsec$), respectively.}
\label{fig:fit_rad}
\end{center}
\end{figure}

To reduce the scatter and highlight radial trends better, we also
applied our two-component model to the radial profiles shown in
Figure~\ref{fig:rprof}, where the radial bins have much higher S/N
than individual pixels. The resulting radial profiles of the age,
stellar mass fraction and surface mass density of the young component
are shown in Figure~\ref{fig:fit_rad}. Errors show the ranges of the
fitting model parameters that are within 1$\sigma$ confidence from the
best-fit model. These clearly show a young stellar component with age
$0.1$--$0.4$~Gyr in the inner $10\arcsec$ and around the outer ring.
The presence of young stars (age $0.2$--$0.4$~Gyr) is also confirmed
at intermediate radii ($10\arcsec\la R\la40\arcsec$). This may seem
somewhat odd considering the FUV map in Figure~\ref{fig:images}, but
it is in agreement with the NUV map. The mass fraction profile in
Figure~\ref{fig:fit_rad} clearly shows that the stellar mass fraction
in the young component is below $1$~per cent
everywhere. Interestingly, the youngest stars are found just inside
$10\arcsec$ (although their mass fraction is very small), where the
[O\,{\small III}] equivalent width map reveals a nuclear ring (see
Section~\ref{sec:rings}). Such rings are usually associated with star
formation, although the H$\beta$/[O\,{\small III}] line ratio in that
case is more characteristic of shocks or low ionization nuclear
emission regions (LINERS; see, e.g., \citealt{setal06}).
%
%

\begin{figure}
\begin{center}
\includegraphics[width=8cm]{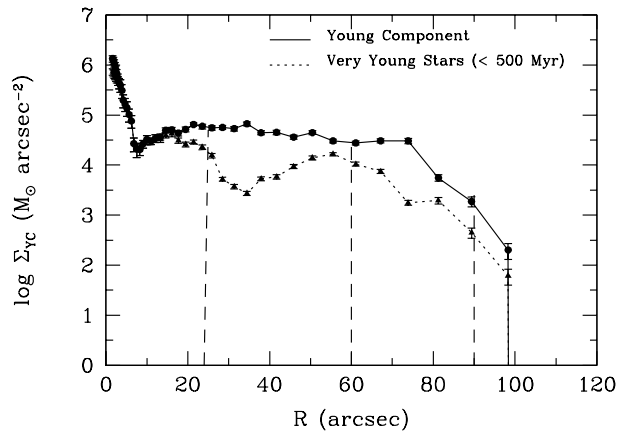}
\caption{Radial profiles of the surface mass density in the young
stellar component, for two different age cuts. While
Figure~\ref{fig:fit_rad}c is based on the radial profiles from ellipse
fitting shown in Figure~\ref{fig:rprof}, this figure is based on
azimuthal averages of the pixel-by-pixel analysis of
Figure~\ref{fig:fit_maps}. The errors are the standard deviation of
the mass fractions derived for individual pixels and should not be
compared to those in Figure~\ref{fig:fit_rad}c. The vertical lines
show the radial positions of the large-scale bar
($R\approx25\arcsec$), the outer ring ($R\approx60\arcsec$) and the
partial ring ($R\approx90\arcsec$), respectively.}
\label{fig:mass}
\end{center}
\end{figure}

The errors bars in Figure~\ref{fig:fit_rad} are driven primarily by
the inability of the stellar population models to properly fit the
data (see Fig.~\ref{fig:fit_ex}), rather than by the uncertainty in
the data themselves. To reduce the scatter in
Figure~\ref{fig:fit_rad}, we have thus azimuthally averaged the
pixel-by-pixel analysis shown in Figure~\ref{fig:fit_maps} along the
same ellipses. The resulting radial profiles of the surface mass
densities in the young component and in``very young'' stars
($<500$~Myr) are shown in Figure~\ref{fig:mass} (see also
Table~\ref{tab:rprof}). The very young stars clearly reproduce the
preponderance of the central regions and outer ring. The total stellar
mass implied by the luminosity of NGC~2974 is about
$1.2\times10^{11}$~M$_{\odot}$, but the total mass in the young
component is only roughly $7.9\times10^{8}$~M$_{\odot}$ or $0.66$~per
cent of the total stellar mass.

The young component dominates the total UV flux of this galaxy but has
a minor ($<10$~per cent) effect in the optical
($\approx5000$~\AA). For example, the outer and partial rings are
hardly detected at $V$ band (see Fig.\ref{fig:images}). Hence, optical
age estimates on this (e.g.\ \citealt{dttft05}; Kuntschner et al.\
2007, in prep.) or any other galaxies with minor RSF would not be
influenced much.

All our modeling results slightly depend on the choice of the input
parameters for the two-component modeling. For instance, if we assume
$14$~Gyr for the age of the old component instead of $12$~Gyr, the
total mass fraction of the young component becomes slightly less,
$0.6$~per cent. No difference is seen for an old component of
$10$~Gyr. This is because at large ages, old populations develop hot
HB stars (UV upturn phenomenon). Since the UV upturn is thought to be
prominent only for large ages, the degeneracy between the UV upturn
and the RSF is negligible for younger ages, i.e.\ $8$--$12$~Gyr. The
choice of metallicity for the young component has no significant
effect on the derived parameters.
%
%

\section{FIGURE ROTATION}
\label{sec:fig_rot}
\subsection{Resonance rings}
\label{sec:rings}
Rings such as those observed in the UV in NGC~2974 are often seen in
disc galaxies, where they can generally be explained by the so-called
resonance ring formation, relating the shapes and positions of the
rings to the presence of resonances in the discs
\citep[e.g.][]{s81,s84a,brsbc94}. Given a circular velocity curve and
the presence of an $m=2$ perturbation such as a bar, the pattern speed
fixes the existence and position of resonances and thus the orbit
families present in the disc (see, e.g., \citealt{sw93}).

\begin{figure}
\begin{center}
\includegraphics[width=8cm]{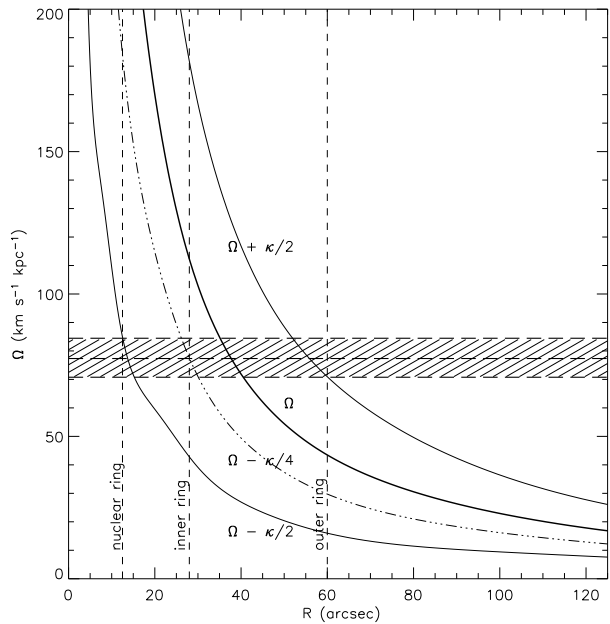}
\caption{Resonance diagram for NGC~2974. Profiles of $\Omega$ (thick
  solid line), $\Omega\pm\kappa/2$ (thin solid lines) and
  $\Omega-\kappa/4$ (dot-dashed line) are shown. Vertical dashed lines
  show the positions of the {\it GALEX} and {\tt SAURON} rings, while
  thick horizontal dashed lines show the implied pattern speeds.}
\label{fig:freq}
\end{center}
\end{figure}

\citet{kcemz05} constructed a detailed mass model of NGC~2974 from
optical observations, in order to reproduce its stellar and gas
kinematics within the radial range probed by their {\tt SAURON}
observations. The mass model nevertheless extends farther and allows
to predict the circular velocity curve (and epicyclic frequency) to
the larger radii probed by {\it GALEX}, where the rotation curve is
currently unknown \citep[but
see][]{kjgkg88,zetal96,wkomz06}. Figure~\ref{fig:freq} shows the
rotation and other frequencies ($\Omega\pm\kappa/2$,
$\Omega-\kappa/4$) predicted from this model, as well as the location
of the major rings observed in the {\it GALEX} UV and {\tt SAURON}
ionised-gas data.
%
%

As discussed in Section~\ref{sec:results}, the UV ring detected in the
{\it GALEX} data has a radius of $60\arcsec$. Placing this on the
frequencies diagram, and assuming that this ring traces the outer
Lindblad resonance (OLR; i.e.\ that the ring is an outer ring), we
obtain a pattern speed of about $72$ km~s$^{-1}$~kpc$^{-1}$ (at the
intersection with the $\Omega+\kappa/2$ curve).

\begin{figure}
\begin{center}
\includegraphics[width=8cm]{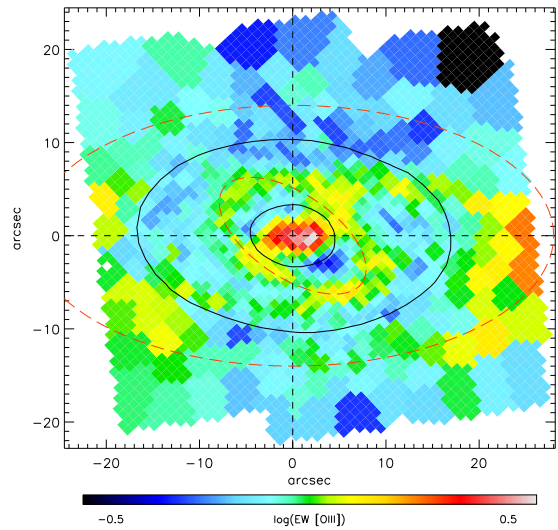}
\caption{Equivalent width map of the [O$\,${\small III}] emission line
  as observed by {\tt SAURON} (adapted from \citealt{kcemz05}).
  Overplotted are two isophotes from the reconstructed broadband image
  (solid lines) and the outlines of the nuclear and inner rings
  discussed in Section~\ref{sec:rings} (dashed lines). The image has
  been rotated so that the galaxy major-axis lies horizontal.}
\label{fig:oiii}
\end{center}
\end{figure}

Although not as clear as the {\it GALEX} UV ring, the [O$\,${\small
III}] equivalent width map of \citeauthor{kcemz05}
(\citeyear{kcemz05}; see also \citealt{setal06}), adapted in
Figure~\ref{fig:oiii}, suggests a (pseudo-)ring near the edge of the
{\tt SAURON} FOV, at a radius of roughly $28\arcsec$. Placing this
ring on the frequencies diagram, and assuming that it traces
corotation (i.e.\ that it is an inner ring), we obtain a pattern speed
of roughly $112$ km~s$^{-1}$~kpc$^{-1}$ (at the intersection with the
$\Omega$ curve). This is inconsistent with the value obtained from the
{\it GALEX} outer ring, even allowing for a large uncertainty in the
radius. However, assuming instead that this inner ring is located at
the inner ultra harmonic resonance (IUHR), as is argued equally often
\citep[e.g.][]{s84b,b95,psa03}, we obtain a pattern speed of roughly
$77$ km~s$^{-1}$~kpc$^{-1}$ (at the intersection with the
$\Omega-\kappa/4$ curve), in good agreement with the value from the
{\it GALEX} outer ring.
%
%

The {\tt SAURON} [O$\,${\small III}] equivalent width map further
reveals a much smaller ring, with a radius of $9\arcsec$ along its
major-axis. This ring is clearly misaligned from the main body of the
galaxy, however, and assuming a galaxy inclination of $59\degr$ (the
value preferred by \citealt{kcemz05}) and an orientation relative to
the galaxy major-axis of $34\degr$, its deprojected radius is
$13\arcsec$. Placing this on the frequencies diagram, and assuming
that this innermost ring traces the inner Lindblad resonance (ILR;
i.e.\ that it is a nuclear ring), we obtain a pattern speed of about
$84$ km~s$^{-1}$~kpc$^{-1}$ (at the intersection with the
$\Omega-\kappa/2$ curve). This value is slightly higher than that
obtained from the {\it GALEX} outer ring and the {\tt SAURON}
[O$\,${\small III}] inner ring, but it is consistent given the
steepness of the $\Omega-\kappa/2$ curve in the inner regions and the
uncertainty on the ring's radius (an increase of only $1$--$2\arcsec$
in the deprojected ring radius brings the pattern speeds in
agreement).
\subsection{Pattern speed}
\label{sec:pat_speed}
Given the above results, we argue that NGC~2974 has a non-negligible
asymmetry and harbours a large-scale $m=2$ pattern, most likely a
large-scale bar, with a pattern speed of $78\pm6$
km~s$^{-1}$~kpc$^{-1}$. The small surface brightness plateau observed
at radii $20$--$25\arcsec$ (see Section~\ref{sec:results}) suggests a
bar length of $\approx23\arcsec$ (end of the plateau), just consistent
with \citet{kcemz05} who suggest a large-scale bar radius of
$12.5$--$25\arcsec$. The bar would then fit nicely inside the
[O\,{\small III}] inner ring (radius $28\arcsec$) but would end well
inside its own corotation radius (see Fig.~\ref{fig:freq}), with a
corotation to bar radius ratio of about $1.6$, slightly higher than
expected and typically observed in barred disc galaxies \citep*[$1.2$
for fast bars; see, e.g.,][]{a92,adc03,gkm03}.  \citet{kcemz05} do
suggest a pattern speed of $185$ km~s$^{-1}$~kpc$^{-1}$, significantly
different from ours, but we are confident in our estimate as higher
values do not allow any major resonance at the large radius where the
{\it GALEX} ring is found (see Fig.~\ref{fig:freq}). For a pattern
speed of $185$ km~s$^{-1}$~kpc$^{-1}$ and a bar radius of 23$\arcsec$,
the bar would then also extend beyond its own corotation radius (see
Fig.~\ref{fig:freq}), a situation forbidden on orbital grounds
\citep[e.g.][]{c80}.

Table~\ref{tab:speeds} presents a compilation of all direct bar
pattern speed measurements in lenticular galaxies, all of them
obtained with the method of \citet{tw84} or variations of it
\citep[e.g.][]{mk95}. As can be seen from Table~\ref{tab:speeds}, all
measured values lie in the range $25$--$105$ km~s$^{-1}$~kpc$^{-1}$. A
value of $78\pm6$ km~s$^{-1}$~kpc$^{-1}$ thus appears normal, while
$185$ km~s$^{-1}$~kpc$^{-1}$ would be exceptional.
%
%

\begin{table}
\caption{Pattern speeds of lenticular galaxies}
\label{tab:speeds}
\begin{tabular}{@{}llrr}
\hline
Galaxy & Type & Pattern speed & Reference\\
& & (km~s$^{-1}$~kpc$^{-1}$) & \\
\hline
NGC~936 & SB(rs)0$^+$ & $104$ & 1\\
NGC~936 & SB(rs)0$^+$ & $60\pm14$ & 2\\
NGC~1023 & SB(rs)0$^-$ & $104\pm34$ & 3\\
NGC~1308 & SB(r)0/a & $99.4\pm34.8$ & 4\\
NGC~1358 & SAB(r)0/a & $31\pm15$ & 5\\
NGC~1440 & (L)SB(rs)0$^+$ & $83.0\pm19.1$ & 4\\
NGC~2950 & (R)SB(r)0$^0$ & $99.2\pm21.2$ & 6\\
NGC~3412 & SB(s)0$^0$ & $56.7\pm15.5$ & 4\\
NGC~4596 & SB(r)0$^+$ & $52\pm13$ & 7\\
IC~874 & SB(rs)0$^0$ & $41.6\pm14.3$ & 4\\
ESO~139-G009 & (RL)SB(rl)0$^0$ & $61.4\pm16.6$ & 4\\
ESO~281-G031 & SB(rl)0$^0$ & $27\pm11$ & 5\\
NGC~7079 & (L)SB(r)0$^0$ & $45.5\pm1.1$ & 8\\
\hline
\end{tabular}

References: (1) \citet{k87}; (2) \citet{mk95}; (3) \citet{dca02}; (4)
\citet{adc03}; (5) \citet{gkm03}; (6) \citet{cda03}; (7)
\citet{gkm99}; (8) \citet{dw04}.
\end{table}
We also note that the pattern speed of \citet{kcemz05} was obtained by
combining the relatively simple gas model of \citet{egf03}, yielding
the pattern speed of the nuclear bar detected, with the assumption
that the corotation of the nuclear bar coincides with the ILR of the
large-scale bar (see Fig.~7 of \citealt{kcemz05}). However, if the
pattern speed of the nuclear bar is accurate, the arguments above
suggest instead that it is the OLR of the nuclear bar (rather than its
corotation) which coincides with the ILR of the large-scale bar,
consistent with the apparent elongation of the nuclear ring
perpendicular to the nuclear bar (see Fig.~\ref{fig:oiii}). Of course,
all of the above arguments assume linear perturbations and localised
resonances, so perfect matches should not be expected, and many
possibilities exist for nuclear bar--large-scale bar (non-linear)
coupling \citep[e.g.][]{fm93,mt97}.

In the [O$\,${\small III}] equivalent width map, the nuclear (radius
$13\arcsec$) and inner (radius $28\arcsec$) rings appear connected by
two arc-like structures reminiscent of spiral arms. The
H$\alpha$+[N$\,${\small II}] map of \citet{egf03} reveals smaller
spiral arms within the nuclear ring, but the two spiral systems appear
to wind in opposite directions. Based on the central dust distribution
\citep[more prominent to the East; e.g.][]{setal06} and the ionised
gas and stellar kinematics \citep[North-East half receding;
e.g.][]{eetal04}, it is clear that the nuclear spiral (within the
nuclear ring) is trailing while the larger-scale spiral (between the
nuclear and inner rings) is leading. To our knowledge, this is the
first time this configuration is observed, but it is unclear how to
generate such a system within the framework of linear perturbation
theory. The reverse combination of a leading nuclear spiral and
trailing larger-scale spiral can be generated by a potential with a
large-scale bar and constant density core \citep[producing an inner
ILR; see, e.g.,][]{w94,w01,m04}, but this is ruled out by
\citeauthor{kcemz05}'s (\citeyear{kcemz05}) mass model and
Figure~\ref{fig:freq} (the potential is cuspy to the smallest
observable scales). We have thus as yet no proper explanation for this
behaviour, but it may well offer a way to better understand the
coupling between nuclear and large-scale bars.
\subsection{Star formation -- dynamics connection}
\label{sec:sf-dyn}
\citeauthor{wkomz06} (\citeyear{wkomz06}; see also Weijmans et al.\
2007, in prep.) have obtained new H\,{\small I} synthesis observations
of NGC~2974, which reveal that the H\,{\small I} is confined to a very
regular although lopsided (i.e.\ more H\,{\small I} to the North-East)
broad ring, extending from roughly $R=50\arcsec$ to
$120\arcsec$. These observations naturally solve the issue of the fuel
for the RSF detected here. The newly-identified UV outer ring at
$60\arcsec$ is located at the inner edge of the H\,{\small I} ring,
while the incomplete UV ring at larger radii is closer to the outer
edge of the H\,{\small I}. Furthermore, the H\,{\small I} lopsidedness
to the North-east matches very well the extra RSF detected there.

Interestingly, while the UV (which is primarily sensitive to young
stars up to $\approx1$~Gyr) does detect a young population, optical
studies (which are primarily sensitive to slightly older stellar
populations; e.g.\ \citealt{dttft05}; Kuntschner et al.\ 2007, in
prep.) have not provided any evidence for young stars. This suggests
either that bursts of star formation previous to the ones detected
here were too small to be detected optically, or that this is the
first burst in a long time. Although the morphology and kinematics of
the H\,{\small I} ring detected by Weijmans et al.\ are very regular,
the lower spatial resolution observations of \citet{kjgkg88} do reveal
a more irregular structure, with a small extension to the
South-East. It is thus possible that the H\,{\small I} was only
recently accreted.

It is unclear whether the ionised gas detected in the inner parts in
the optical (e.g.\ Fig.~\ref{fig:oiii}) originates in the H\,{\small
I}. Bar torques are such that gas normally flows from corotation to
the OLR, not the other way around. This can explain why the H\,{\small
I} has accumulated in the ring observed by Weijmans et al., but makes
it difficult for any gas to trickle down the center. Whether the
ionised gas has the same origin as the H\,{\small I} thus presumably
depends on exactly how the H\,{\small I} was accreted.
%
%
\section{SUMMARY}
\label{sec:summary}
We have presented {\it GALEX} FUV and NUV and ground-based F555W
imaging of the nearby early-type galaxy NGC~2974. The optical images
are characteristic of elliptical galaxies but the UV images reveal
both a central spheroid-like component and a complete (outer) ring of
radius $6.2$~kpc. Another partial ring is also suggested at an even
larger radius. Both rings are detected here for the first time and are
only visible in the UV. The FUV luminosity of the outer ring is
$\approx1.2~\times10^{8}$~$L_{\odot}$.

Blue FUV$-$NUV colours are observed in the centre of the galaxy and
around the outer ring, suggesting the presence of young stellar
populations ($\la1$~Gyr) and recent star formation. UV$-$optical
colours are also blue at both locations, indicating that (very) young
stars contribute significantly to the galaxy's UV flux. Simple
two-component stellar population modeling allows us to constrain the
age, stellar mass fraction and surface mass density of the young
component pixel by pixel and confirms this behaviour. The latter are
formally lower limits since we have not corrected for dust extinction,
but the effect is expected to be small.

Considering the mass model and observations of \citet{kcemz05}, the
{\tt SAURON} [O$\,${\small III}] nuclear and inner rings and the {\it
GALEX} UV outer ring are all consistent with a single pattern speed of
$78\pm6$ km~s$^{-1}$~kpc$^{-1}$, suggesting that NGC~2974 harbours a
large-scale $m=2$ asymmetry such as a bar. Assuming that the surface
brightness plateau observed at a radius of $20$--$25\arcsec$ marks its
limits, this large-scale bar would end at roughly $60$~per cent of its
own corotation radius, slightly smaller but not inconsistent with
expectations.

Star formation, rings and bars are not topics normally associated with
elliptical galaxies, yet in recent years we have seen a growing number
of studies reporting evidence for complex kinematics
\citep[e.g.][]{eetal04} and recent star formation
\citep[e.g.][]{yetal05}. The origin of those features are of key
importance. Here, we have witnessed evidence for recent star formation
in a particular early-type galaxy, in the form of a ring, fueled by a
newly discovered co-spatial ring of neutral hydrogen
\citep{wkomz06}. A rotating bar (i.e.\ a disc phenomenon) seems to be
regulating both phenomena, adding to earlier evidence for
lenticular-like properties in NGC~2974. Star formation may thus not be
all that unusual in early-type galaxies, or perhaps elliptical
galaxies in the classical sense are simply far rarer than usually
assumed.
%
%
\section*{Acknowledgments}
The authors thank the referee, S.\ Trager, for numerous comments
which led to improvement in the paper. We would like to thank S.\ Kaviraj
and K.\ Schawinski for useful discussions, and T.\ de Zeeuw,
E.\ Emsellem, H.\ Kuntschner, R.\ Peletier and M.\ Sarzi for comments
on the manuscript.  HJ acknowledges support from the International
Research Internship Program of the Korea Science and Engineering
Foundation. MB acknowledges support from NASA through GALEX Guest
Investigator program GALEXGI04-0000-0109. This work was supported by grant
No.~R01-2006-000-10716-0 from the Basic Research Program of the Korea
Science and Engineering Foundation to SKY. RLD is grateful for the
award of a PPARC Senior Fellowship (PPA/Y/S/1999/00854) and
postdoctoral support through PPARC grant PPA/G/S/2000/00729. The PPARC
Visitors grant (PPA/V/S/2002/00553) to Oxford also supported this
work. Based on observations made with the NASA Galaxy Evolution
Explorer. GALEX is operated for NASA by the California Institute of
Technology under NASA contract NAS5-98034. Photometric data were also
obtained using the 1.3m McGraw-Hill Telescope of the MDM
Observatory. Part of this work is based on data obtained from the
ESO/ST-ECF Science Archive Facility. This project made use of the
HyperLeda and NED databases.
%
%

%
\end{document}